\documentclass[aip,jcp,amsmath,amssymb,reprint,floatfix]{revtex4-1}

\usepackage{graphicx}
\usepackage{dcolumn}
\usepackage{bm}
\usepackage{color}
\usepackage{hyperref}
\usepackage{multirow}

\newcolumntype{d}[1]{D{.}{.}{#1}}

\usepackage[utf8]{inputenc}
\usepackage[T1]{fontenc}
\usepackage{mathptmx}

\newcommand{\half}{\frac{1}{2}}

\newcommand{\vecR}{\mathbf{R}}

\begin{document}

\preprint{AIP/123-QED}

\title{Determination of fundamental properties of nitrogen from first principles. \\ III. Temperature and frequency dependence of the molecular polarizability and magnetic susceptibility}

\author{Jakub Lang}
\affiliation{University of Warsaw, Faculty of Chemistry, Pasteura 1, 02-093 Warsaw, Poland}
\author{Giovanni Garberoglio}
\affiliation{European Centre for Theoretical Studies in Nuclear Physics and Related Areas (FBK-ECT*), Strada delle Tabarelle 286, 38123 Trento, Italy}
\author{Micha\l\ Przybytek}
\author{Micha\l\ Lesiuk}
\email{e-mail: m.lesiuk@uw.edu.pl}
\affiliation{University of Warsaw, Faculty of Chemistry, Pasteura 1, 02-093 Warsaw, Poland}
\date{\today}

\begin{abstract}
This work is the third part of the series of papers that focus on the theoretical determination of the properties of nitrogen that are relevant in metrology. Here we present first-principles calculations of the temperature and frequency dependence of the molecular polarizability and magnetic susceptibility of the nitrogen molecule (N$_2$). The purely electronic contributions to the static polarizability, Cauchy coefficients (up to sixth order), and isotropic magnetic susceptibility are computed over a range of internuclear distances using a robust composite scheme combining several electronic structure methods. The temperature dependence, evaluated from $50$~K to $2000$~K, is determined using two independent methods: rovibrational averaging and path integral Monte Carlo (PIMC). The polarizabilities obtained from theory agree with the recent high-precision thermometry measurements, wherever the latter are available, but are significantly less accurate. However, the main usefulness of the theoretical data revolves around combining it with the available experimental results to generate semi-empirical estimates of various quantities that have never been measured thus far. As an example, we determine highly accurate semi-empirical estimates of the static polarizability at key reference temperatures, $\alpha_0(T)=11.735\,962$~a.u.\ at $T=303$~K and $\alpha_0(T)=11.735\,585$~a.u.\ at $T=273.16$~K. Furthermore, we report theoretical values for the magnetic susceptibility, highlighting the importance of the paramagnetic contribution, and address a significant discrepancy with the experimental data for this quantity.
\end{abstract}

\maketitle

\section{Introduction}
\label{sec:intro}

In the first two papers of the series, we have determined the interaction potential and other quantities that describe the electronic ground state of the N$_2$ molecule directly from theory. This enables us to focus now on the properties of this molecule that are directly relevant from the point of view of modern thermometry and precision metrology in general. \cite{garberoglio23} For example, in the case of refractive index gas thermometry (RIGT), \cite{jousten17,gao17,rourke19,ripa21,rourke21} the link between the measured macroscopic quantities and microscopic properties of the molecule that can be obtained from the first principles is established using the Lorentz--Lorenz equation: \cite{lorentz80,lorenz80}
\begin{align}
\label{ll}
    \frac{n^2(\omega,T)-1}{n^2(\omega,T)+2} = \frac{4\pi N_A}{3} \left[ \alpha(\omega,T) + \chi(\omega,T) \right]\rho,
\end{align}
where $n(\omega,T)$ is the refractive index of the gas that depends on the angular frequency $\omega$ of the resonant electromagnetic field and temperature $T$. In the above expression $N_A$ is the Avogadro constant, $\rho$ is the molar density of the gas, while $\alpha(\omega,T)$ and $\chi(\omega,T)$ are the dynamic polarizability and magnetic susceptibility of the N$_2$ molecule. Note that the above equation holds only approximately, i.e., for sufficiently low densities of the gas such that the interaction between N$_2$ molecules can be neglected. In a separate future work, we will report the calculation of the refractive virial coefficients which account for corrections due to interaction-induced polarization.

The application of the Lorentz--Lorenz equation to mol\-ecules is complicated by the fact that the polarizability and magnetic susceptibility depend both on the frequency and temperature. Their temperature dependence originates from the fact that each rovibrational state of the molecule has a slightly different expectation value of the polarizability or magnetic susceptibility when averaged over the corresponding rovibrational wavefunction. As the occupation of these states varies with temperature (according to the Boltzmann distribution in the classical limit), this dependence propagates to the quantities observed in the experiments. Note that for monoatomic gases, such as noble gases, this effect is entirely absent and hence the overall temperature dependence appears only in higher-order terms (virial coefficients). Therefore, measurement of, e.g. the refractive index, for a diatomic molecule within the whole range of admissible temperatures and frequencies is time consuming or even impossible due to the limitation of experimental setups, and hence theoretical calculations represent an attractive alternative to a large number of independent measurements. Such a hybrid approach enables us to use measured data for an isolated set of combinations of $\omega$ and $T$ which are available, and supplement it by the frequency and temperature dependence obtained from theory to shift it to any other value of these parameters that is required in a particular application. However, we are not aware of any theoretical work in the literature where the temperature and frequency dependence of the polarizability and magnetic susceptibility of nitrogen molecule have been studied in a systematic way. Therefore, the main purpose of this work is to fill this gap.

To determine $\alpha(\omega,T)$ and $\chi(\omega,T)$ directly from theory, some additional considerations are required. First, calculations of such properties are usually not performed directly at a specified temperature. Instead, polarizability and magnetic susceptibility are determined as a function of the internuclear distance $R$ in the N$_2$ molecule, which is written as $\alpha(\omega,R)$ and $\chi(\omega,R)$. To translate these quantities into $\alpha(\omega,T)$ and $\chi(\omega,T)$, we adopt two alternative approaches in the present work: averaging over rovibrational levels with weights obtained from the canonical ensemble, and the path integral Monte Carlo (PIMC) approach. \cite{Feynman65,Ceperley95,Garberoglio24} Both methods take into account the quantum effects beyond the classical or semiclassical picture and should yield, in principle, the same results within reasonable numerical limits. Therefore, the use of two alternative approaches enables us to validate the results and check for potential errors.

The second consideration is related to the frequency dependence of the quantities in question. In principle, one could calculate $\alpha(\omega,R)$ and $\chi(\omega,R)$ for a set of optical frequencies that are relevant to RIGT experiments. However, an interpolation would be required to calculate them in-between the adopted frequency grid if such a need arises in the future. A much more convenient approach is to express, e.g., the dynamic polarizability, as a power series of the angular frequency:
\begin{align}
\label{cauchyexp}
    \alpha(\omega,R) = \alpha_0(R) + \alpha_2(R)\,\omega^2 + 
    \alpha_4(R)\,\omega^4 + \alpha_6(R)\,\omega^6 + \ldots,
\end{align}
where $\alpha_0(R)$ is the static polarizability (the $\omega\rightarrow0$ limit) while $\alpha_n(R)$ are the so-called Cauchy coefficients. Note that the same expansion was used in Paper~I in the case of the nitrogen atom. For N$_2$, the range of applicability of this expansion is somewhat smaller due to the decreased first resonance frequency of the molecule, but still covers the optical range of $\omega$ relevant to RIGT measurements (wavelengths longer than $400\,$nm or $\omega<0.1$ in atomic units) and is rapidly convergent there. Therefore, in this work, we report the values of $\alpha_n(R)$ up to $n=6$ rather than the polarizabilities at a set of finite frequencies.

The following conversion factors are used throughout the present work: $1$~hartree = $27.211\,386$~eV = $219\,474.63$~cm$^{-1}$ for energy, and $1$~bohr = $0.529\,177\,2$~\AA{} for length, according to the most recent 2022 CODATA recommendations. \cite{mohr25} The conversion factor for polarizability and magnetic susceptibility is $1$~cm$^3$/mol = $2.675\,205$~a.u. and for angular frequency $1$~a.u. = $4.134\,137\,336 \times 10^{16}\,$rad/s.

\section{Polarizability of a diatomic molecule}
\label{sec:polar}

Rigorous theoretical determination of the polarizability of a diatomic molecule is a delicate matter, because it calls for treatment of the vibrational and rotational degrees of freedom in addition to the electronic ones. In exact theory, all these degrees of freedom are coupled, even in the Born--Oppenheimer approximation, \cite{born1927quantentheorie,kutzelnigg1997adiabatic} which prevents us from finding a formula that takes them into account separately. As the determination of the polarizability from such a coupled electronic-nuclear formalism is computationally intractable, one usually adopts additional approximations that force separation of the electronic and nuclear motions, as discussed in the review article of Bishop. \cite{bishop90} Typically, one assumes that the rovibrational energy spacings are small in comparison with the electronic ones and neglects the former in parts of the expressions. Although such an approximation is physically justified for a molecule such as N$_2$, it is difficult to assess its accuracy. Therefore, here we provide a different derivation of the expression for the polarizability of a homonuclear diatomic molecule. It yields the same leading-order expression as in Ref.~\onlinecite{bishop90}, but enables us to find the formula for the dominant correction and assess its magnitude without resorting to heuristic arguments.

We consider a homonuclear diatomic molecule in a weak external electric field within the Born--Oppenheimer approximation. The center of mass motion is separated from the total Hamiltonian and does not couple to the field; hence it will be neglected in the following. The electronic wavefunctions $|n\rangle$ are eigenvectors of the purely electronic Hamiltonian $H_\mathrm{el}$ in relative coordinates with the eigenvalues $E_n(R)$. Both quantities carry a parametric dependence on the internuclear distance $R$. For the wavefunctions $|n\rangle$, this dependence is omitted in the notation for clarity. The rovibrational wavefunctions $|\nu_n J_n\rangle$ are solutions of the Schr\"odinger equation for the relative motion of the nuclei:
\begin{align}
    \left( T + E_n(R) \right) |\nu_n J_n\rangle = E_{n\nu_n J_n} |\nu_n J_n\rangle,
\end{align}
where $T=-\frac{1}{2\mu}\nabla_R^2$ is the kinetic energy operator for the relative nuclear motion ($\mu$ is the reduced mass of the molecule). Further on, we will use the concise notation $H_n = T + E_n(R)$. To account for the interaction with the external electric field, we adopt the minimal coupling Hamiltonian in the dipole gauge, in which the electronic Hamiltonian acquires the term $-\sum_i \mathbf{r}_i\cdot \mathbf{E}$, where the index $i$ denotes electrons and $\mathbf{E}$ is the electric field oscillating with frequency $\omega$.

Let us assume that the molecule is initially in a definite quantum state $|0\nu_0 J_0\rangle$ prior to the application of the field. To avoid notational clutter, we will write $|0\nu J\rangle\equiv |0\nu_0 J_0\rangle$ in the following. Note that the initial electronic state is always assumed to be the ground state. The expression for the dynamic polarizability of a molecule in this state is obtained using the standard second-order perturbation theory, taking into account all degrees of freedom. We are interested only in the isotropic polarizability $\alpha_{\nu J}(\omega)$ averaged over the orientations of the molecule, as this quantity is relevant for gas thermometry experiments.
It takes the form:
\begin{align}
\begin{split}
    \alpha_{\nu J}(\omega) &= \frac{1}{3}
    \sum_{n\neq 0}\sum_{\nu_nJ_n} \bigg[
    \frac{|\langle 0\nu J|\mathbf{d}|n\nu_nJ_n\rangle|^2}{E_{n\nu_nJ_n} - E_{0\nu J}-\omega} \\
    &+ \frac{|\langle 0\nu J|\mathbf{d}|n\nu_nJ_n\rangle|^2}{E_{n\nu_nJ_n} - E_{0\nu J}+\omega}
    \bigg],
\end{split}
\end{align}
where $\mathbf{d}=-\sum_i \mathbf{r}_i$ is the total electronic dipole operator. Note that for a homonuclear diatomic molecule we have $\langle 0|\mathbf{d}|0\rangle=0$ which allowed us to eliminate some of the cross-terms in the above expression.

First, we expand the dynamic polarizability in powers of $\omega$ to isolate the static polarizability and Cauchy coefficients. This leads to:
\begin{align}
    \alpha_{\nu J}(\omega) = \sum_{k=0}^\infty \alpha_{\nu J}^{(2k)}\,\omega^{2k}.
\end{align}
As discussed in the introduction, this expansion is entirely sufficient in the optical and microwave regime. The coefficients take the form:
\begin{align}
    \alpha_{\nu J}^{(2k)} = \frac{2}{3}
    \sum_{n\neq 0}\sum_{\nu_nJ_n} 
    \frac{|\langle 0\nu J|\mathbf{d}|n\nu_nJ_n\rangle|^2}{(E_{n\nu_nJ_n} - E_{0\nu J})^{2k+1}}.
\end{align}
In the Born--Oppenheimer approximation, the wavefunctions $|n\nu_nJ_n\rangle$ are products of the electronic and rovibrational components, i.e., $|n\nu_nJ_n\rangle\equiv|n\rangle|\nu_nJ_n\rangle$, and hence we can integrate over the electronic degrees of freedom in the above formula:
\begin{align}
    \alpha_{\nu J}^{(2k)} = \frac{2}{3}
    \sum_{n\neq 0}\sum_{\nu_nJ_n} 
    \frac{|\langle \nu J|\mathbf{d}_{0n}(R)|\nu_nJ_n\rangle|^2}{(E_{n\nu_nJ_n} - E_{0\nu J})^{2k+1}},
\end{align}
where $\mathbf{d}_{0n}(R)=\langle 0|\mathbf{d}|n\rangle$ is the transition dipole moment which depends parametrically on $R$. Because the quantities $|\nu_nJ_n\rangle$ are eigenfunctions of the Hamiltonian $H_n$, the equivalent form of the above equation reads:
\begin{align}
\begin{split}
    \alpha_{\nu J}^{(2k)} &= \frac{2}{3}
    \sum_{n\neq 0}\sum_{\nu_nJ_n} 
    \langle \nu J|\mathbf{d}_{0n}(R)|\nu_nJ_n\rangle \\
    &\times\langle \nu_nJ_n| (H_n - E_{0\nu J})^{-2k-1}\,\mathbf{d}_{0n}(R)|\nu J\rangle.
\end{split}
\end{align}
We now use the closure relation for the rovibrational wavefunctions $|\nu_nJ_n\rangle$ to eliminate the explicit sum over these states:
\begin{align}
\label{boaa}
\begin{split}
    \alpha_{\nu J}^{(2k)} &= \frac{2}{3}
    \sum_{n\neq 0}
    \langle \nu J|\mathbf{d}_{0n}(R) (H_n - E_{0\nu J})^{-2k-1}\,\mathbf{d}_{0n}(R)|\nu J\rangle.
\end{split}
\end{align}
Using the properties of the Hamiltonian $H_n$, we can write $H_n=H_0 + E_n(R) - E_0(R)$. The usual approximation applied here is to neglect the kinetic energy operator in the above formula due to its inverse dependence on the reduced mass, as well as the term $E_{0\nu J}$ which is small compared to the electronic energy differences $E_n(R) - E_0(R)$. This leads to the standard formula:
\begin{align}
\label{viban}
\begin{split}
    \alpha_{\nu J}^{(2k)} &= \frac{2}{3}
    \sum_{n\neq 0}
    \frac{\langle \nu J|\mathbf{d}_{0n}(R) \mathbf{d}_{0n}(R)|\nu J\rangle}{(E_n(R) - E_0(R))^{2k+1}},
\end{split}
\end{align}
see Ref.~\onlinecite{bishop90}, and after application of the closure relation for the sum over the electronic states $|n\rangle$ we find:
\begin{align}
\label{alpha0nuj}
    \alpha_{\nu J}^{(2k)} &= \langle \nu J|\alpha_{2k}(R)|\nu J\rangle,
\end{align}
where the term in the brackets is the pure electronic contribution to the isotropic Cauchy coefficient:
\begin{align}
\label{alpha0nuj2}
    \alpha_{2k}(R) &= \frac{2}{3}
    \langle0| \mathbf{d}\,Q\big( H - E_0 \big)^{-2k-1}Q\,\mathbf{d} |0\rangle,
\end{align}
where $Q=1-|0\rangle\langle0|$ is the projector onto the subspace orthogonal to the electronic ground state.

To derive the above expression, we have neglected the kinetic energy operator in Eq.~(\ref{boaa}), an approximation which is uncontrolled at the moment. To judge the quality of this approximation, we have to derive an expression for the leading-order correction and estimate its magnitude. For technical simplicity, we focus on the static polarizability ($k=0$), but the derivation proceeds almost the same for the Cauchy coefficients. The expansion of the resolvent operator can be generalized to the form:
\begin{align}
\begin{split}
    &(H_n - E_{0\nu J})^{-1} = (H_0 + E_n(R) - E_0(R) - E_{0\nu J})^{-1} \\
    &=(E_n(R) - E_0(R))^{-1} \\
    &- (E_n(R) - E_0(R))^{-1} (H_0 - E_{0\nu J})(E_n(R) - E_0(R))^{-1} \\
    &+ \ldots
\end{split}
\end{align}
The first term on the right-hand side is exactly the denominator in Eq.~(\ref{viban}), so retaining only this term leads directly to Eq.~(\ref{viban}). The second term on the right-hand side constitutes the leading-order correction not accounted for in Eq.~(\ref{viban}). Inserting this formula into Eq.~(\ref{boaa}) with $k=0$ we obtain the corresponding correction to the static polarizability:
\begin{align}
\begin{split}
    \Delta\alpha_{\nu J}^{(0)} &= -\frac{2}{3}
    \sum_{n\neq 0}
    \langle \nu J|\mathbf{F}_{0n}(R) (H_0 - E_{0\nu J})\,\mathbf{F}_{0n}(R)|\nu J\rangle,
\end{split}
\end{align}
where $\mathbf{F}_{0n}(R) = \mathbf{d}_{0n}(R)(E_n(R) - E_0(R))^{-1}$. Using the fact that $(H_0 - E_{0\nu J})|\nu J\rangle=0$, we can rewrite the above formula in the commutator form:
\begin{align}
\begin{split}
    \Delta\alpha_{\nu J}^{(0)} &= -\frac{1}{3}
    \sum_{n\neq 0}
    \langle \nu J|\Big[ \mathbf{F}_{0n}(R) \big[H_0,\mathbf{F}_{0n}(R)\big]\Big]|\nu J\rangle  \\
    &=\frac{1}{6\mu}
    \sum_{n\neq 0}
    \langle \nu J|\Big[ \mathbf{F}_{0n}(R) \big[\nabla_R^2,\mathbf{F}_{0n}(R)\big]\Big]|\nu J\rangle.
\end{split}
\end{align}
As the electronic potential $E_0(R)$ and the quantity $\mathbf{F}_{0n}(R)$ are both multiplicative functions of $R$, only the kinetic energy term $T$ from the total Hamiltonian $H_0=T + E_0(R)$ survives in the commutator, hence the second equality. Writing $\nabla_R^2$ in the spherical coordinate system, using the fact that $\mathbf{F}_{0n}(R)$ depends only on $R$ as the sole variable, and taking the derivatives one arrives at:
\begin{align}
\label{corrpol}
    \Delta\alpha_{\nu J}^{(0)} = -\frac{1}{3\mu}\sum_{n\neq 0}
    \langle \nu J|\partial_R \mathbf{F}_{0n}(R)\cdot \partial_R \mathbf{F}_{0n}(R)|\nu J\rangle.
\end{align}
Due to the presence of the inverse reduced mass in the above formula (for $^{14}$N$_2$ $\mu\approx 12\,800$ in atomic units), this term is expected to be small, but it is difficult to judge \emph{a priori} the value of the unknown term in the brackets. Therefore, we perform an estimation of the $\Delta\alpha_{\nu J}^{(0)}$ correction by adopting a series of simplifications. First, the dominant contribution to the temperature dependence of the polarizability comes from the $\nu=0$ and $J=0$ term, so it is reasonable to focus only on this particular case:
\begin{align}
\label{corrpol0}
    \Delta\alpha_{00}^{(0)} = -\frac{1}{3\mu}\sum_{n\neq 0}
    \langle 00|\partial_R \mathbf{F}_{0n}(R)\cdot \partial_R \mathbf{F}_{0n}(R)|00\rangle.
\end{align}
Next, to obtain a rough estimate of the value of the above expression, we consider the first dipole-allowed excited electronic state, which should bring the dominant contribution to the sum over $n$. States that are dipole-forbidden are guaranteed to have zero contribution due to the presence of the transition dipole moment $\mathbf{d}_{0n}(R)$ in the quantity $\mathbf{F}_{0n}(R)$. As the ground electronic state of the N$_2$ molecule is characterized by the molecular term $X^1\Sigma_g^+$, only $^1\Sigma_u^+$ and $^1\Pi_u$ states have to be taken into account, and the lowest-energy dipole-allowed state turns out to be $b^1\Pi_u$ in the spectroscopic notation. The potential energy curve $E_n(R)$ and the transition dipole moment $\mathbf{d}_{0n}(R)$ for this state were taken from the work of Hammami \emph{et al.} \cite{hammami26} The calculated raw data for a set of internuclear distances $R$ were interpolated by us using the third-order $B$-splines, and the functions $\mathbf{F}_{0n}(R)$ and $\partial_R \mathbf{F}_{0n}(R)$ are evaluated analytically based on this interpolation. Finally, integration over $R$ in Eq.~(\ref{corrpol0}) that includes the rovibrational wavefunction $|00\rangle$ is carried out as described later in Sec.~\ref{sec:temp}. We found that the $\Delta\alpha_{00}^{(0)}$ correction is equal to roughly $-2\cdot 10^{-4}$ in atomic units which is entirely negligible from the point of view of this work as the total polarizability is five orders of magnitude larger. Of course, the correction term in Eq.~(\ref{corrpol0}) would increase if a larger number of excited states were included, but even an increase of $\Delta\alpha_{00}^{(0)}$ by a factor of $50$, which is extremely unlikely, would not change the overall conclusion. 

\section{Temperature dependence}
\label{sec:temp}

As mentioned in the introduction, two independent methods are adopted in this work for determination of the temperature dependence of the reported quantities. In this section we introduce the basics of both methods and perform some estimates regarding the range of internuclear distances for which the target quantities need to be found. Our target temperature range is $50-2000\,$K which is entirely sufficient from the point of view of gas thermometry.

\subsection{Rovibrational averaging}
\label{sec:rovibro}

Let $\mathcal{O}(R)$ be any quantity of interest, e.g., Cauchy coefficient or magnetic susceptibility, known as a function of the internuclear distance $R$. To determine the temperature dependence of this quantity, we first average $\mathcal{O}(R)$ over a set of rovibrational wavefunctions $|\nu J\rangle\equiv\chi_{\nu J}(R)$ according to the formula:
\begin{align}
\label{onuj}
    \mathcal{O}_{\nu J} = \int_0^\infty \mathrm{d}R\; R^2\,|\chi_{\nu J}(R)|^2\,\mathcal{O}(R).
\end{align}
In the case of the static polarizability and Cauchy coefficients, the origin of this formula was explained in Sec.~\ref{sec:polar}, see Eqs.~(\ref{alpha0nuj})~and~(\ref{alpha0nuj2}), but the results translate without changes to any second-order quantity $\mathcal{O}_{\nu J}$, hence a more general notation is used here. We do not assume the separation of vibrational and rotational degrees of freedom, i.e., the nuclear Schr\"odinger equation for the relative motion of the nuclei is solved for each $J$ separately including the explicit centrifugal term $J(J+1)/R^2$.

The temperature dependence of the quantity $\mathcal{O}$ is found by averaging over all admissible rovibrational levels:
\begin{align}
\label{oaver}
    \mathcal{O} = \sum_{\nu J} \mathcal{O}_{\nu J}\,P_{\nu J}(T),
\end{align}
where the weight $P_{\nu J}(T)$ represents the probability that the state $\nu J$ is occupied at the given temperature $T$. These probabilities are obtained within the canonical ensemble:
\begin{align}
    P_{\nu J}(T) = g_{\nu J}\,\frac{e^{-E_{\nu J}/kT}}{\mathcal{Q}(T)},
\end{align}
where $g_{\nu J}$ is the degeneracy of the level $\nu J$, while the partition function $\mathcal{Q}(T)$ ensures that the probabilities are normalized to unity at any temperature:
\begin{align}
    \mathcal{Q}(T) = \sum_{\nu J} g_{\nu J}\,e^{-E_{\nu J}/kT}.
    \label{eq:Q}
\end{align}
Note that due to the normalization of $P_{\nu J}(T)$, the choice of the zero energy reference level does not matter; for convenience and numerical stability, we set $E_{00}=0$ in contrast to Paper~II where the rovibrational energies were calculated with respect to the bottom of the interaction potential well. Additionally, it is worth pointing out that the use of the canonical ensemble to calculate the probabilities is justified because even for the highest temperature considered here ($T=2000\,$K), bond breaking and formation in N$_2$ practically does not occur. Consequently, the number of molecules present in a given volume remains constant. In fact, the dissociation energy of N$_2$ is roughly $D_0 = 9.76\,$eV, see Paper~II, and hence this assumption is justified even for several times larger temperatures and becomes violated to a significant degree only at around $10^4\,$K which is irrelevant from the point of view of thermometry.

In calculation of the averages, we have to take into account the spin of the nitrogen nucleus. In this analysis, we consider the $^{14}$N--$^{14}$N isotopologue as the most commonly used in thermometry. The $^{14}$N nucleus is a boson with spin quantum number $I=1$. Therefore, there are three possible spin functions for a single nucleus or nine possible spin functions for the $^{14}$N--$^{14}$N dimer. For temperatures considered here ($T>50\,$K), these spin functions are practically degenerate and have essentially identical occupations. Out of the nine possible spin functions, six are symmetric with respect to the exchange of the nuclei in the diatomic molecule, while three are antisymmetric. As the ground $X^1\Sigma_g^+$ electronic state is symmetric with respect to this operation, for the six nuclear spin functions which are symmetric with respect to the exchange of the nuclei only even values of $J$ are allowed, while for the remaining three only odd values of $J$ are allowed. To take this into account we define the degeneracies $g_{\nu J}$ according to the formula:
\begin{align}
    g_{\nu J} = 
    \begin{cases}
    6(2J+1) \mbox{\;\;for even $J$}\\
    3(2J+1) \mbox{\;\;for odd $J$}
    \end{cases},
\end{align}
where the factor $2J+1$ is the same as in the rigid rotor approximation.

In calculation of the rovibrational averages, there is no practical need to perform the summation in Eq.~(\ref{oaver}) over all states supported by the potential. Indeed, the contributions to Eq.~(\ref{oaver}) vanish very quickly with increasing $\nu J$, because of the exponential factor $e^{-E_{\nu J}/kT}$ present in $P_{\nu J}(T)$. This enables us to select the maximum energy (and hence the number) of rovibrational levels needed in the averaging procedure at the given level of uncertainty. To this end, note that if $P_{\nu J}(T)$ is below $10^{-4}$ then such level can be neglected, because the error of the order of $10^{-4}$ is negligible in comparison with the overall uncertainty of the theoretical data reported in this work, as well as in Papers~I and II. Even for the largest temperature of interest, $T=2000\,$K, the value of $P_{\nu J}(T)=10^{-4}$ corresponds to the maximum energy of the state $E_{\nu J}$ equal to roughly $12\,800\,$cm$^{-1}$ (calculated with respect to $E_{00}=0$). With the rovibrational levels found in Paper~II we see that, for example, for $J=0$ we have to take into account the first six vibrational levels $\nu=0,1,\ldots,5$. In total, there are 345 rovibrational levels with energies less than $12\,800\,$cm$^{-1}$ above the vibrational zero point energy, and we take all such combinations of $\nu J$ into account when evaluating the temperature dependence of any quantity.

This truncation of the number of rovibrational levels leads to another simplification. Note that the rovibrational levels with energies below $12\,800\,$cm$^{-1}$ are concentrated around the bottom of the potential energy well. They do not extend too far to the asymptotic region or do not penetrate the repulsive wall of the potential in the sense that their amplitudes are non-negligible only in a relatively small vicinity of the equilibrium distance, $R_e$. Therefore, the integrals $\mathcal{O}_{\nu J}$ defined in Eq.~(\ref{onuj}) can be truncated as follows:
\begin{align}
\label{onuj2}
    \mathcal{O}_{\nu J} \approx \int_{R_-}^{R_+} \mathrm{d}R\; R^2\,|\chi_{\nu J}(R)|^2\,\mathcal{O}(R),
\end{align}
without substantial deterioration of the quality of the results, because the probability amplitude $|\chi_{\nu J}(R)|^2$ is effectively zero outside the integration interval $(R_-,R_+)$. The major simplification achieved through this truncation is the fact that the quantity $\mathcal{O}(R)$ does not have to be determined over the same range of $R$ as the interaction potential considered in Paper~II. Instead, the knowledge of this quantity $\mathcal{O}(R)$ only within the interval $(R_-,R_+)$ is entirely sufficient. This is an important point, because calculation of the polarizability or Cauchy coefficients of the nitrogen molecule over the whole range of $R$ would be extremely difficult due to the multireference nature of this system in the bond-breaking regions.

To exploit this simplification, we have to determine the values of $R_-$ and $R_+$ in Eq.~(\ref{onuj2}) such that the error resulting from the truncation is acceptably small. To this end, we look at the magnitude of $R^2\,|\chi_{\nu J}(R)|^2$ for the 345 rovibrational levels necessary for the averaging, as a function of $R$. By using the data from Paper~II, we found that for $R_-=1.75$~bohr and $R_+=2.60$~bohr, the magnitude of $R^2\,|\chi_{\nu J}(R)|^2$ outside the interval $(R_-,R_+)$ is below $10^{-6}$. Therefore, truncation of the integral in Eq.~(\ref{onuj2}) to the interval $(R_-,R_+)$ is expected to bring an error of less than one part per million in determination of $\mathcal{O}_{\nu J}$. At a preliminary stage of the project, we additionally tested this finding by calculating the averages for some model functions $\mathcal{O}(R)$ representing the polarizability and progressively increasing the range of $(R_-,R_+)$. We never encountered any change in the final temperature dependence that would exceed $10^{-6}$ and hence the adopted range $(R_-,R_+)$ is a safe choice from the practical point of view.

\subsection{Path integral Monte Carlo}
\label{sec:pimc}

\begin{figure}
\includegraphics[width=0.95\linewidth]{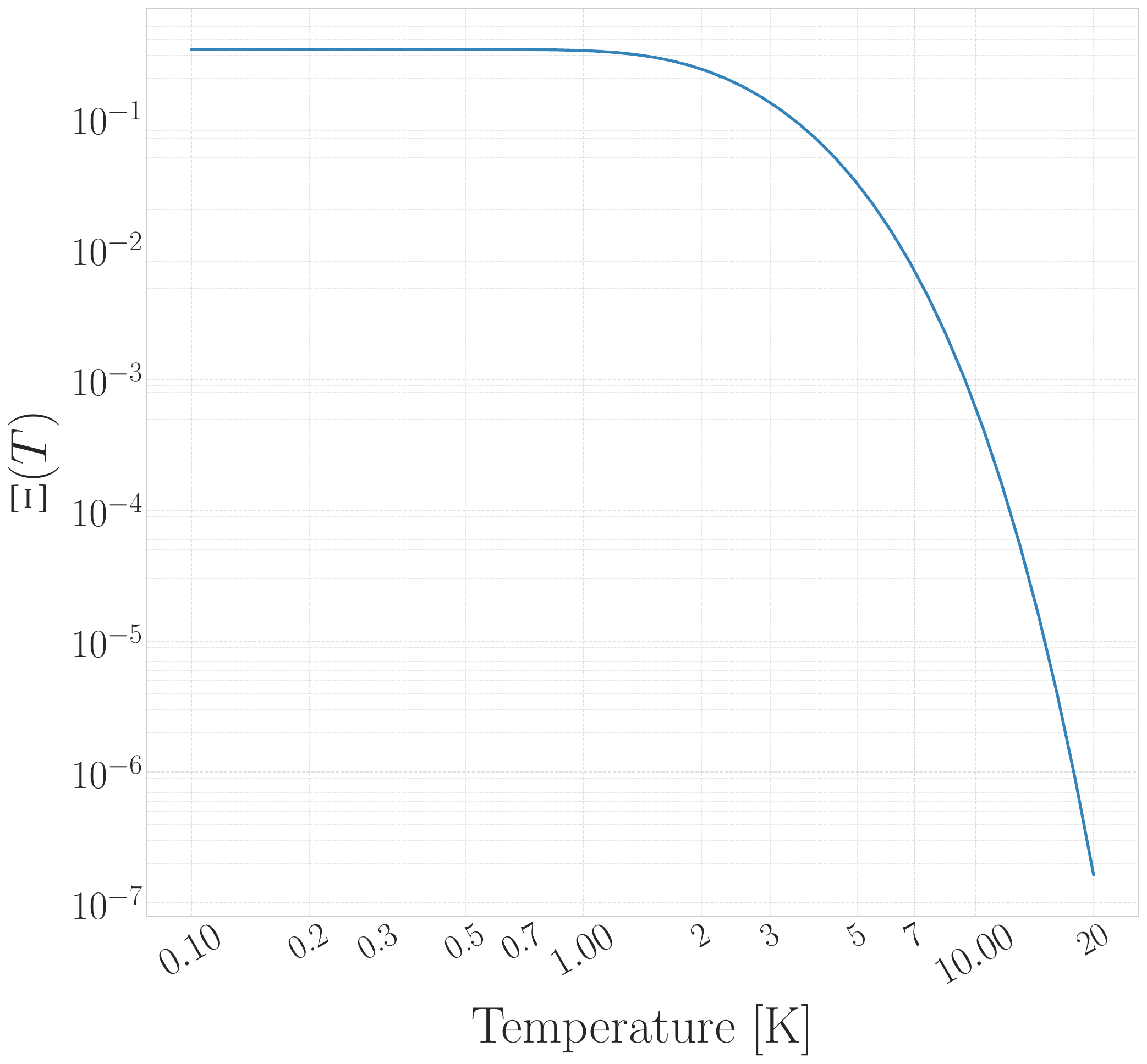}
\caption{
Temperature dependence of the partition function ratio $\Xi(T) = {\cal Q}_\mathrm{xc}(T) / {\cal Q}_\mathrm{B}(T)$ for the $\text{N}_2$ molecule, where ${\cal Q}_\mathrm{xc}$ and ${\cal Q}_\mathrm{B}$ are the exchange [Eq. (\ref{eq:QB})] and Boltzmann [Eq. (\ref{eq:Qxc})] contributions to the partition function, respectively. 
}
\label{fig:Xi}
\end{figure}

In the PIMC method, \cite{Feynman65,Ceperley95} the partition function (\ref{eq:Q}) for a homonuclear diatomic molecule is written as ${\cal Q} = ({\cal Q}_\mathrm{B} + {\cal Q}_\mathrm{xc})/2$, where
\begin{eqnarray}
    {\cal Q}_\mathrm{B}(T) &=& \mathrm{tr}\left(e^{-\beta H}\right) \nonumber\\
    &=& (2I+1)^2 \int \langle \vecR^{(1)} | e^{-\beta H} | \vecR^{(1)} \rangle \;\mathrm{d}^3\vecR^{(1)}, \label{eq:QB}\\
    {\cal Q}_\mathrm{xc}(T) &=& \mathrm{tr}\left(e^{-\beta H} {\cal P}\right) \nonumber \\
    &=& (2I+1) \int \langle \vecR^{(1)} | e^{-\beta H} | -\vecR^{(1)} \rangle 
    \;\mathrm{d}^3 \vecR^{(1)},
    \label{eq:Qxc}
\end{eqnarray}
where $H = T + V$ is the Hamiltonian corresponding to the relative coordinate $\vecR^{(1)}$ [a subscript (1) has been introduced for later convenience], $V = E_0(R)$ is the interaction potential between the two N atoms in their ground state, and the trace operator is taken over {\em all} quantum states. The operator $\cal P$ exchanges the two nuclei, and would include a factor of $-1$ if they were fermions, which is not the case for ${}^{14}$N. ${\cal Q}_\mathrm{B}$ is called the Boltzmann term (since it corresponds to {\em distinguishable} particles), while ${\cal Q}_\mathrm{xc}$ is called the exchange term. These two terms have very different temperature behaviours. In particular, the exchange term is usually only relevant at small temperatures. \cite{Garberoglio14,Garberoglio18} The values of the ratio $\Xi(T) = {\cal Q}_\mathrm{xc}(T) / {\cal Q}_\mathrm{B}(T)$ is shown in Fig.~\ref{fig:Xi}, where one can see that the exchange term is completely negligible for $T \gtrsim 10$~K. Since in this work we will be concerned with much higher temperatures, we will disregard ${\cal Q}_\mathrm{xc}(T)$ in the following.

In this case, \cite{Garberoglio24} the PIMC method proceeds by using the Trotter factorization, i.e., writing $e^{-\beta H} = \left( e^{-\beta H / P} \right)^P$ in Eq.~(\ref{eq:QB}) and using the high-temperature expansion $e^{-\beta H / P} \sim e^{-\beta T / P} e^{-\beta V / P}$. By further using $P-1$ completeness relation between the various exponentials---hence introducing the variables $\vecR^{(k)}$ for $k=2, \ldots, P$---one arrives at the well-known classical ring-polymer isomorphism that maps the Boltzmann partition function of two quantum particles to a classical partition function of a ring polymer with $P$ monomers. \cite{Garberoglio09} The mapping is exact in the $P \to \infty$ limit. \cite{Feynman65}
In the PIMC approach, the equivalent of Eq.~(\ref{oaver}) is
\begin{equation}
    {\cal O} = \frac{1}{{\cal Q}_\mathrm{B}}\int \left\langle 
    \overline{\cal O} e^{-\beta \overline{V}} 
    \right\rangle \mathrm{d}^3 \vecR^{(1)},
    \label{eq:Oavg_PIMC}
\end{equation}
where
\begin{eqnarray}
    \overline{\cal O} &=& \frac{1}{P} \sum_{k=1}^P {\cal O}\left(\vecR^{(k)}\right), \label{eq:Obar}\\
    \overline{V} &=& \frac{1}{P} \sum_{k=1}^P V\left(\vecR^{(k)}\right),
    \label{eq:Vbar}
\end{eqnarray}
are the average values of the observable $\cal O$ and the interaction potential $V$ on the positions of all the $P$ monomers of the classical ring-polymer.
The brackets in Eq.~(\ref{eq:Oavg_PIMC}) denote the average over the possible configurations of a classical ring polymer, whose probability distribution---specified by the quantum-classical mapping---is that of a classical Brownian bridge starting and ending at $\vecR^{(1)}$. Notice that the probability distribution for the Brownian bridge, used to calculate the bracket average in Eq.~(\ref{eq:Oavg_PIMC}), depends on the relative distances between monomers $\vecR^{(k+1)}-\vecR^{(k)}$, with $\vecR^{(P+1)} \equiv \vecR^{(1)}$.

In actual applications, the value of $P$ must be chosen large enough to guarantee convergence of the observable averages in Eq.~(\ref{eq:Oavg_PIMC}) to the actual value within a specified accuracy. In our case, we found that we could converge the PIMC averages to value much smaller than the propagated uncertainty (to be discussed below) by using $P = \mathrm{nint}\left(42\,000~\mathrm{K}/T\right)$ where $\mathrm{nint}(x)$ denotes the integer closest to $x$. We computed the various observables by generating $10^7$ ring polymer configurations via Metropolis sampling and using one every ten to perform the averages, in order to guarantee independent samples. We used the Levy construction \cite{Levy54,Fosdick66} to generate trial moves by resampling $p$ consecutive  monomer positions, adjusting $p$ to obtain an acceptance ratio $\approx 50\%$.

We used functional differentiation of Eq.~(\ref{eq:Oavg_PIMC}) to evaluate how the uncertainty on $\cal O$ and $V$ propagates to the uncertainty on the average values. \cite{Garberoglio21,Wheatley23,Garberoglio26}
The propagation of the uncertainty on $\cal O$ is straightforward, since
\begin{equation}
     \delta {\cal O}_{\cal O} = \frac{1}{{\cal Q}_\mathrm{B}} \int \left\langle \overline{\delta {\cal O}} 
     e^{-\beta \overline{V}} 
    \right\rangle \mathrm{d}^3 \vecR^{(1)},
    \label{eq:unc_O}
\end{equation}
where $\overline{\delta {\cal O}}$ is the average of the uncertainty over all the relative coordinates $\vecR^{(k)}$, analogously to Eqs.~(\ref{eq:Obar}) and (\ref{eq:Vbar}) above.

The contribution of the uncertainty $\delta V$ of the pair potential to the uncertainty of the observable $\cal O$ can be obtained using the same approach, remembering that the Boltzmann partition function also depends on $V$. After some lengthy but straightforward calculations, one obtains
\begin{eqnarray}
\left| \delta{\cal O}_V \right|&=& \left| \frac{1}{{\cal Q}_\mathrm{B}} \int \left\langle 
    \overline{\cal O} ~ \beta \overline{\delta V} e^{-\beta \overline{V}}
 \right\rangle \mathrm{d}^3 \vecR^{(1)} -   \right.\nonumber \\
 & & \left. \int 
    \frac{\left\langle 
    \overline{\cal O} e^{-\beta \overline{V}}
    \right\rangle}{{\cal Q}_\mathrm{B}}   \;\mathrm{d}^3 \vecR^{(1)}
  \int 
    \frac{\left\langle 
    \beta \overline{\delta V} e^{-\beta \overline{V}}
    \right\rangle}{{\cal Q}_\mathrm{B}}   \;\mathrm{d}^3 \vecR^{(1)}\right| \nonumber \\
    &\equiv& \left|\mathrm{Cov}\left(\overline{\cal O}, \beta\overline{\delta V}\right)\right|,
    \label{eq:unc_V}
\end{eqnarray}
where in the last line we have recognized the expression that comes from functional differentiation as the covariance of the quantities $\overline{\cal O}$ and $\beta\overline{\delta V}$.

Unfortunately, this expression is not completely satisfactory, since the covariance of a pair of variables is not guaranteed to be positive definite. In fact, it might happen that for some values of the temperature the covariance crosses zero, implying the unphysical result of zero propagated uncertainty. Notice that this cannot happen in the propagation of the uncertainty of the observable $\cal O$ in Eq.~(\ref{eq:unc_O}), since the right-hand side is positive definite.
In order to overcome the limitation of Eq.~(\ref{eq:unc_V}), we propose to use the Cauchy--Schwartz inequality $\left|\mathrm{Cov}(\overline{\cal O}, \beta\overline{\delta V})\right| < 
\sqrt{\mathrm{Var}(\overline{\cal O})\mathrm{Var}(\beta \overline{\delta V}) }$ and estimate the uncertainty propagated from the potential as the upper bound
\begin{equation*}
    \delta{\cal O}_V = \sqrt{\mathrm{Var}(\overline{\cal O}) ~ \mathrm{Var}(\beta \overline{\delta V})},
    \label{eq:unc_V_CS}
\end{equation*}
which is positive definite.
The averages needed to compute $\delta {\cal O}_{\cal O}$ and $\delta{\cal O}_V$ were computed during the same production runs described above.
Finally, the two uncertainties have been combined in quadrature to obtain the overall uncertainty
\begin{equation}
    \delta{\cal O} = \sqrt{\delta{\cal O}_{\cal O}^2 + \delta{\cal O}_V^2}.
    \label{eq:U_total}
\end{equation}
We consider the final uncertainties as being expanded uncertainties with coverage factor $k=2$, corresponding approximately to a 95\% confidence level. We found $\delta{\cal O}_V \ll \delta{\cal O}_{\cal O}$ at all the temperatures investigated in this work.

\section{Ab initio calculations}
\label{sec:abinitio}

\subsection{Polarizability and Cauchy coefficients}
\label{subsec:alpha}

In this section, we focus on the \emph{ab initio} calculation of the purely electronic contribution to the (isotropic) static polarizability and Cauchy coefficients, defined in Eq.~(\ref{alpha0nuj2}), for the nitrogen molecule for internuclear distances within the range $(R_-,R_+)$. As we are dealing with regions close to the equilibrium internuclear distance, a composite method based on the coupled-cluster (CC) hierarchy of methods \cite{vcivzek1966correlation,vcivzek1969use,crawford00,bartlett07} is the preferred approach, similarly as for the short-range part of the interaction potential developed in Paper~II. After some experimentation, we adopted the following composite scheme:
\begin{align}
\label{composite}
\begin{split}
    \alpha_n(R) &= 
    \alpha_n^{\mathrm{fc-CCSD}}(R) 
    + \delta \alpha_n^{\mathrm{fc-CC3}}(R) \\
    &+ \delta \alpha_n^{\mathrm{fc-CCSDT}}(R)
    + \delta \alpha_n^{\mathrm{fc-CCSDTQ}}(R) \\
    &+ \delta \alpha_n^{\mathrm{ae-CC3}}(R)
    + \delta \alpha_n^{\mathrm{ae-CCSDT}}(R)
    + \delta \alpha_n^{\mathrm{rel1}}(R),
\end{split}
\end{align}
both for the static polarizability ($n=0$) and for the Cauchy coefficients ($n=2,4$). The naming conventions for the particular corrections are the same as in Paper~II, i.e., the method used to calculate a particular correction (CCSD, \cite{purvis1982full,scuseria1987closed} CC3, \cite{koch97} CCSDT, \cite{noga1987full,scuseria1988new} CCSDTQ \cite{kucharski1991recursive,oliphant1991coupled,kucharski1992coupled}) is given in the superscript. Moreover, the acronym ``fc'' stands for frozen-core (two $1s^2$ orbitals of nitrogen atoms kept uncorrelated), and ``ae'' stands for all-electron, i.e., all electrons are included in the correlated treatment. All calculations were performed using the same family of basis sets, aug-cc-pV$X$Z, \cite{dunning89,kendall92,woon94,wilson96} as in Paper~II. For the internuclear distances considered here, further augmentation of the basis set leads to no appreciable improvement in the quality of the results. The last Cauchy coefficient $\alpha_6(R)$ contributes only at a level of a few parts per million to the polarizability at optical and microwave frequencies and does not have to be calculated as accurately as the other quantities. Therefore, only the first two terms in Eq.~(\ref{composite}) were included in the calculation of $\alpha_6(R)$.

In Paper~II, we reached accuracy slightly better than 0.1\% in the interaction potential near the minimum of the potential energy curve. It is unreasonable to expect that the polarizability and Cauchy coefficients can be calculated with a meaningfully higher accuracy. Therefore, corrections to the polarizability and Cauchy coefficients that contribute less than 0.1\% (combined) will be neglected in this work. This enables us to make some simplifications in the composite scheme used for the calculations of the polarizability, see Eq.~(\ref{composite}). This choice is guided primarily by the magnitude of some terms considered in Paper~II as well as on preliminary calculations of the polarizability. First of all, we entirely neglect here the quantum electrodynamics (QED) \cite{BeSal,Pachucki2004} contributions and the diagonal Born--Oppenheimer correction (DBOC). \cite{born1927quantentheorie,kutzelnigg1997adiabatic} These two terms together contribute less than 0.008\% to the total interaction energy near the minimum of the interaction potential, see Paper~II. Therefore, even if their values grow by an order of magnitude in the case of the polarizability, they are still negligible in comparison with the other sources of error. Moreover, calculations of the polarizability for noble gases available in the literature \cite{lesiuk23,lesiuk20} suggest that there is no reason to suspect an anomalous growth of these corrections in the present case. Second, the two-electron relativistic corrections are also neglected in the present work. In Paper~II, we found that they are at least several times smaller than the one-electron corrections to the interaction energies, and it is likely that the same holds true for the polarizability. This assumption is additionally supported by the data available for neon and argon. As we shall see shortly, the one-electron relativistic correction $\delta \alpha_n^{\mathrm{rel1}}(R)$ contributes no more than 0.1--0.2\% to the static polarizability and Cauchy coefficients, making the two-electron relativistic corrections negligible. The remaining differences between the composite scheme~(\ref{composite}) and the protocol adopted in Paper~II are explained below, where the particular contributions are discussed separately.

The calculations of the first two components, $\alpha_n^{\mathrm{fc-CCSD}}(R)$ and $\delta \alpha_n^{\mathrm{fc-CC3}}(R)$, were performed using the {\sc Dalton} package. The implementation reported in Ref.~\onlinecite{dalton} enables direct calculations of the static polarizability and Cauchy coefficients. \cite{christiansen1998integral,hattig1997cauchy,pawlowski05} Note that, in contrast with Paper~II, the component $\alpha_n^{\mathrm{fc-CCSD}}(R)$ includes also the Hartree--Fock (HF) \cite{Hartree_1928,fock1930naherungsmethode} contribution to the polarizability, which was not treated separately. This is because we observed that the HF contribution converges much faster with the basis set size than the remaining correlated corrections, so with the two largest basis sets used for $\alpha_n^{\mathrm{fc-CCSD}}(R)$, the differences were essentially negligible. Such phenomenon is well-known in the literature \cite{feller92,halkier99,lesiuk2019ab} and also is fully aligned with the findings of Paper~II regarding the HF contribution to the interaction energy, where the rapid convergence of the HF term was additionally confirmed by using numerical grid-based HF calculations. The $\alpha_n^{\mathrm{fc-CCSD}}(R)$ contribution was calculated using aug-cc-pV$X$Z basis sets up to $X=7$, while calculation of $\delta \alpha_n^{\mathrm{fc-CC3}}(R)$ is feasible up to $X=6$. Extrapolation to the complete basis set (CBS) limit and estimation of the residual basis set incompleteness error are performed using the same protocol as in Papers~I and II, i.e., the Riemann extrapolation formula \cite{lesiuk19Riemann} followed by the random walk uncertainty estimation scheme. \cite{Lang25random} The same approach is used for all remaining contributions, unless explicitly stated otherwise. As an illustration, in Table~\ref{tab:ansummary} we report the magnitudes of $\alpha_n^{\mathrm{fc-CCSD}}(R)$ and $\delta \alpha_n^{\mathrm{fc-CC3}}(R)$ terms, as well as their respective uncertainties, for three internuclear distances: $R=1.750$~bohr, $R=2.070$~bohr, and $R=2.600$~bohr. These distances correspond to the outer edges of the integration interval $(R_-,R_+)$ used in the rovibrational averaging, see Eq.~(\ref{onuj2}), as well as to the minimum of the interaction potential which usually contributes the most to any temperature-dependent quantity. The corresponding results for the whole grid of internuclear distances considered in this work are given in the Supplementary Material. \cite{supp}

\begin{table*}
\caption{\label{tab:ansummary}
Summary of the results for static polarizability and Cauchy coefficients for internuclear distances $R=1.750$, $R=2.070$, and $R=2.600$. Expanded $(k=2)$ uncertainty estimates of each term and the sum are given in parentheses. All quantities are given in atomic units.
}
\begin{ruledtabular}
\begin{tabular}{ld{10}d{10}d{10}}
 contribution & 
 \multicolumn{1}{c}{$R=1.750$} & 
 \multicolumn{1}{c}{$R=2.070$} & 
 \multicolumn{1}{c}{$R=2.600$} \\
 \hline\\[-1.2em]
 & \multicolumn{3}{c}{static polarizability $\alpha_0(R)$} \\
 \hline\\[-1.2em]
$\phantom{\delta}\alpha_0^{\mathrm{fc-CCSD}}(R)$ & 9.8249(54) & 11.5847(68) &  14.572797(25)\\
$\delta \alpha_0^{\mathrm{fc-CC3}}(R)$  & 0.0453(24) & 0.1588(63) &  0.909(19)\\
$\delta \alpha_0^{\mathrm{fc-CCSDT}}(R)$ & -0.0112(42) & -0.0565(52) & -0.4982(32)\\
$\delta \alpha_0^{\mathrm{fc-CCSDTQ}}(R)$ & -0.0063(32) & -0.0099(49) & -0.071(35) \\
$\delta \alpha_0^{\mathrm{ae-CC3}}(R)$ &  -0.02304(48) & -0.03128(22) & -0.04885(49) \\
$\delta \alpha_0^{\mathrm{ae-CCSDT}}(R)$ & -0.00080(15) & -0.00175(36) & -0.0079(20) \\
$\delta \alpha_0^{\mathrm{rel1}}(R)$ &  -0.00330(33)  &  -0.00213(26) &  -0.000218(97) \\
 \hline\\[-1.2em]
total & 9.8255(80) & 11.642(12) & 14.856(40) \\
 \hline\\[-1.2em]
 & \multicolumn{3}{c}{Cauchy coefficient $\alpha_2(R)$} \\
 \hline\\[-1.2em]
$\phantom{\delta}\alpha_2^{\mathrm{fc-CCSD}}(R)$ & 20.7225(44) & 28.045(11) &  43.372(43)\\
$\delta \alpha_2^{\mathrm{fc-CC3}}(R)$  & 0.5263(53) & 1.860(12) & 14.248(25) \\
$\delta \alpha_2^{\mathrm{fc-CCSDT}}(R)$  & -0.031(19) & -0.424(17) &  -7.507(73)\\
$\delta \alpha_2^{\mathrm{fc-CCSDTQ}}(R)$ & 0.000118(59) & 0.131(65) &  0.80(40) \\
$\delta \alpha_2^{\mathrm{ae-CC3}}(R)$ & -0.0734(55) & -0.1063(82) & -0.203(19) \\
$\delta \alpha_2^{\mathrm{ae-CCSDT}}(R)$ &  -0.0069(14) &  -0.0168(34)  &  -0.111(28) \\
$\delta \alpha_2^{\mathrm{rel1}}(R)$ &  -0.0190(19)  &  -0.017363(65)  &  -0.0179(30) \\
 \hline\\[-1.2em]
total & 21.118(21) & 29.471(70)  & 50.58(41) \\
 \hline\\[-1.2em]
 & \multicolumn{3}{c}{Cauchy coefficient $\alpha_4(R)$} \\
 \hline\\[-1.2em]
$\phantom{\delta}\alpha_4^{\mathrm{fc-CCSD}}(R)$ & 58.248(69) & 88.878(57) &  173.732(53)\\
$\delta \alpha_4^{\mathrm{fc-CC3}}(R)$  & 2.476(31)  & 10.436(56) & 128.38(28) \\
$\delta \alpha_4^{\mathrm{fc-CCSDT}}(R)$ & -0.043(68) & -1.71(23) &  -68.08(11) \\
$\delta \alpha_4^{\mathrm{fc-CCSDTQ}}(R)$ & 0.034(17) & 1.02(51) & 10.2(51) \\
$\delta \alpha_4^{\mathrm{ae-CC3}}(R)$ & -0.248(54) & -0.324(53) &  -0.59(11)\\
$\delta \alpha_4^{\mathrm{ae-CCSDT}}(R)$ &  -0.006(28) &  -0.03586(64) &  -0.94(23) \\
$\delta \alpha_4^{\mathrm{rel1}}(R)$ &  -0.04(15) &  -0.167(61)  &  -0.211(41)  \\
 \hline\\[-1.2em]
total & 60.43(19) & 98.09(57) &  242.5(51)\\
\end{tabular}
\end{ruledtabular}
\end{table*}

Before moving on to the remaining corrections, we have to discuss a technical issue related to the calculation of some contributions to the Cauchy coefficients. In the case of $\alpha_n^{\mathrm{fc-CCSD}}(R)$ and $\delta \alpha_n^{\mathrm{fc-CC3}}(R)$, we are able to calculate them directly using the approach from Refs.~\onlinecite{hattig1997cauchy,pawlowski05}. Unfortunately, this is not possible at present for the remaining terms in Eq.~(\ref{composite}) as no implementation of such an approach is available in any computational package known to us. As a result, the contributions to the Cauchy coefficients of the remaining terms in Eq.~(\ref{composite}) have to be calculated indirectly, i.e., by first determining the dynamic polarizability for a finite set of frequencies and subsequently fitting these results with the power expansion analogous to Eq.~(\ref{cauchyexp}). The grid of frequencies used for this purpose comprises 27 points which are listed in the Supplementary Material. \cite{supp} Note that the fitting procedure may potentially introduce a non-negligible error due to the finite size and range of the grid. Fortunately, this error is expected to be largely independent of the level of theory used and hence we can judge its magnitude using the CC3 method. To this end, we calculate the Cauchy coefficients using (i) the fitting approach and (ii) directly as described in the previous paragraph. In the absence of fitting errors, both approaches must give exactly the same results, so any deviations are attributed solely to the fitting error. We found that the fitting procedure introduces essentially no error for the static polarizability, while the second, $a_2(R)$, and fourth, $a_4(R)$, Cauchy coefficients deviate by around 0.1\% and 1\%, respectively. These findings are very weakly dependent on the size of the basis set used in the calculations. Note that the fitting approach is used in this work only for small corrections which themselves constitute no more than 1\% of the total value of the respective Cauchy coefficients. Therefore, the uncertainties resulting from the fitting impact the calculated Cauchy coefficients at the level of 0.01\% or less, which is negligible. Consequently, we do not take this source of uncertainty into account in the overall error budget. The remaining non-relativistic terms were calculated using the MRCC program \cite{mester2025overview,mrcc} interfaced to the CFOUR program package. \cite{matthews2020coupled,cfour}
 
The term $\delta \alpha_n^{\mathrm{fc-CCSDT}}(R)$ takes into account the remaining effects of the triple excitations \cite{kallay2001higher,kallay2004calculation} with respect to the Hartree--Fock determinant which are not reproduced by the approximate CC3 model. They were calculated within aug-cc-pV$X$Z basis sets up to $X=4$, followed by extrapolation to the CBS limit. The results for three representative internuclear distances are shown in Table~\ref{tab:ansummary}. Rather surprisingly, the $\delta \alpha_n^{\mathrm{fc-CCSDT}}(R)$ contributes nearly a quarter of the total triple-excitation contributions to the polarizability and Cauchy coefficients, suggesting that the CC3 model in our particular situation is somewhat less accurate than typically assumed. Taking into account that the term $\delta \alpha_n^{\mathrm{fc-CCSDT}}(R)$ was calculated in a smaller basis set than $\delta \alpha_n^{\mathrm{fc-CC3}}(R)$, they bring a similar contribution to the overall uncertainty despite the latter being several times larger on average. 

Next, we move on to the correction originating from the quadruple excitations, $\delta \alpha_n^{\mathrm{fc-CCSDTQ}}(R)$. Due to the steep computational scaling of the CCSDTQ method, these calculations are feasible only within the smallest aug-cc-pVDZ basis set. However, in the case of the interaction energy, we found this basis to deliver useful results, differing from the best estimates of the CBS limit by less than 10\%. This is consistent with other data available in the literature suggesting that the higher-order correlation effects converge increasingly faster with respect to the basis set size. Nonetheless, we attach a very large uncertainty of 50\% to the $\delta \alpha_n^{\mathrm{fc-CCSDTQ}}(R)$ term calculated in this way, see Table~\ref{tab:ansummary}. These wide error bars are also expected to account for the higher-order correlation effects beyond quadruples. In Paper~II, we found that post-CCSDTQ corrections to the interaction energy are by an order of magnitude smaller than the pure quadruples corrections. Even if they are somewhat larger here, the 50\% uncertainty of the $\delta \alpha_n^{\mathrm{fc-CCSDTQ}}(R)$ term accounts both for these effects and the lack of basis set saturation of the pure quadruples contribution.

All corrections discussed thus far accounted for the correlation effects originating from the valence electrons. As in Paper~II, the effect of core $1s^2$ orbitals of nitrogen is expected to be non-negligible. The dominant correction in Eq.~(\ref{composite}) that takes this into account is $\delta \alpha_n^{\mathrm{ae-CC3}}(R)$ calculated at the CC3 level of theory with all electrons correlated with basis sets up to $X=4$. The data shown in Table~\ref{tab:ansummary} reveals that $\delta \alpha_n^{\mathrm{ae-CC3}}(R)$ constitutes roughly 0.3\% of the total static polarizability and Cauchy coefficients, and hence it is not negligible in the present context. This prompted us to investigate the impact of higher-order correlation effects originating from the core electrons. In contrast with the valence contributions, the influence of the post-CC3 effects in the core correlation is marginal. This can be seen from the values of $\delta \alpha_n^{\mathrm{ae-CCSDT}}(R)$ which were calculated using the complete CCSDT model with basis sets up to $X=3$ (and extrapolated to the CBS limit). The correction $\delta \alpha_n^{\mathrm{ae-CCSDT}}(R)$ is by an order of magnitude smaller than $\delta \alpha_n^{\mathrm{ae-CC3}}(R)$ and hence well below the 0.1\% mark in all cases. For this reason, we neglected the higher-order corrections such as $\delta \alpha_n^{\mathrm{ae-CCSDTQ}}(R)$ which are absent in Eq.~(\ref{composite}). Calculations were conducted using aug-cc-pCVXZ basis sets. \cite{woon1995gaussian}

Finally, we consider the one-electron relativistic corrections to the polarizability. Taking them into account in the same way as in Paper~II, i.e., using perturbation theory based on the Breit--Pauli Hamiltonian, is computationally challenging. Therefore, we follow the approach from Paper~I based on the spin-free second-order Douglas--Kroll--Hess (DKH2) method. \cite{Douglas:1974,Hess:1986,Jansen:1989} As discussed in Paper~I, both approaches differ by terms of order $1/c^4$ which are negligible for light systems, so the results should be essentially equivalent; simultaneously, the DKH2 approach is technically less complicated and requires only straightforward modification of the one-electron Hamiltonian. 
The correction $\delta \alpha_n^{\mathrm{rel1}}(R)$ was calculated using the all-electron CCSD method within decontracted aug-cc-pV$X$Z basis sets up to $X=4$ with the {\sc NWChem} program package \cite{nwchem} and extrapolated to the CBS limit. The mixed relativistic and higher-order correlation effect is clearly negligible, and the use of a larger basis is unnecessary due to relatively small magnitude of this term. 
In Table~\ref{tab:ansummary} we report calculated values of $\delta \alpha_n^{\mathrm{rel1}}(R)$ for three representative internuclear distance. Interestingly, while the importance (in relative terms) of the relativistic correction weakly depends on the internuclear distance, it increases fast with $n$, i.e., the order of the Cauchy coefficient. 
For example, at $R=2.070$ the relativistic correction is essentially negligible for the static polarizability ($\approx0.02\%$), but its contribution grows to $\approx0.06\%$ for the second Cauchy coefficient, and to $\approx~0.17\%$ for the fourth. A similar effect was observed for the atomic polarizabilities in Paper~I.

It is also interesting to discuss the leading sources of uncertainty in calculation of the first three Cauchy coefficients. In the case of the static polarizability, all four valence correlation contributions bring a comparable contribution to the overall uncertainty budget. In other words, reducing just one of them would not increase the accuracy of the final results in a meaningful way. However, this changes when moving to the higher-order Cauchy coefficients. The uncertainty of $\alpha_2(R)$ is dominated to a large degree by the uncertainty of the $\delta \alpha_2^{\mathrm{fc-CCSDTQ}}(R)$ term. In the case of $\alpha_4(R)$, the uncertainty of $\delta \alpha_4^{\mathrm{fc-CCSDTQ}}(R)$ is also dominant, but the $\delta \alpha_4^{\mathrm{fc-CCSDT}}(R)$ term also contributes significantly to the total uncertainty. Clearly, corrections accounting for higher-order coupled cluster excitations limit the accuracy of the $\alpha_2(R)$ and $\alpha_4(R)$ coefficients, while for $\alpha_0(R)$ all valence contributions have a similar effect on the total uncertainty. Finally, the uncertainty of the core-valence corrections and of the relativistic contributions is almost negligible for the first three Cauchy coefficients.

We reiterate that the last Cauchy coefficient, namely $\alpha_6(R)$, is calculated using a simplified composite approach in which only the first two terms in Eq.~(\ref{composite}) are retained, but the remaining details of the computational protocol are identical to those for the lower-order coefficients. To take into account the missing corrections, we attach a flat 10\% uncertainty to the values of $\alpha_6(R)$ obtained in this way, irrespective of the nuclear distance. This estimate is based on the observation that in the case of the $\alpha_4(R)$ coefficient, all terms beyond the first two in Eq.~(\ref{composite}) contribute less than 2\% to the total value. However, as some of these terms increase with the order $n$ of the Cauchy coefficient, we conservatively increased the uncertainty fivefold. The final values of the $\alpha_6(R)$ coefficient are given in the Supplementary Material for the whole grid of internuclear distances. \cite{supp} Finally, in the Supplementary Material we provide plots illustrating the dependence of the static polarizability and the Cauchy coefficients as a function of the internuclear distance. In the case of the static polarizability, this dependence is close to linear within the range of $R$ considered here. However, for the Cauchy coefficients $\alpha_n(R)$, $n=2,4,6$, the corresponding curves become convex and this effect becomes more pronounced with increasing $n$.

\subsection{Magnetic susceptibility}
\label{subsec:chi0}

The magnetic susceptibility of an atomic or molecular system is defined as the second derivative of the energy with respect to the strength of the uniform magnetic field $\mathbf{B}=(B_x,B_y,B_z)$ in the zero-field limit $|\mathbf{B}|\rightarrow 0$. As such, it is in principle a symmetric tensor with six independent components:
\begin{align}
    \chi_{\mu\nu} = -\left(\frac{\partial^2 E}{\partial B_\mu \partial B_\nu}\right)\bigg|_{|\mathbf{B}|=0},
\end{align}
where $\mu,\nu=x,y,z$. However, in most experiments the molecules in a sample are either randomly oriented or allowed to rotate freely, which makes only the isotropic part, i.e., the trace of the tensor, observable:
\begin{align}
    \chi_0 = -\frac{1}{3}\left( 
    \frac{\partial^2 E}{\partial B_x^2} + \frac{\partial^2 E}{\partial B_y^2} + \frac{\partial^2 E}{\partial B_z^2}
    \right)\bigg|_{|\mathbf{B}|=0}.
\end{align}
Consequently, the calculation of this quantity is our main focus here.

The semi-classical treatment of the magnetic susceptibility for light atoms, i.e., without quantizatization of the photonic degrees of freedom, is based on the theory of Van Vleck. \cite{van1928dielectric} Here we give only the key formulas that are relevant for a diatomic molecule. We start with the minimal coupling electronic Hamiltonian for the molecule in the uniform magnetic field $\mathbf{B}$ after separation of the center-of-mass motion and dropping all terms proportional to the inverse of the nuclear mass (Born--Oppenheimer approximation):
\begin{align}
\label{hel}
    H_{\mathrm{el}} = \sum_i \left( \mathbf{p}_i + \frac{1}{c}\mathbf{A}_i \right)^2 + V,
\end{align}
where the index $i$ denotes electrons, $\mathbf{p}_i$ is the momentum operator for $i$-th electron, $\mathbf{A}_i = \half\left(\mathbf{B}\times \mathbf{r}_i\right)$ is the vector potential at the position of the electron, and $V$ is the electron-electron and electron-nuclei interaction potential which is not affected by the presence of the field. The coordinates of all particles are measured with respect to their common center of mass. Assuming the Coulomb gauge, $\mathbf{p}_i\cdot \mathbf{A}_i = 0$, and expanding the above expression into powers of $\mathbf{A}_i$ we arrive at:
\begin{align}
    H_{\mathrm{el}} = H_0 + H_1 + H_2,
\end{align}
where $H_0$ is the Hamiltonian for a molecule without the field, while $H_1$ and $H_2$ gather terms linear and quadratic in the field, respectively:
\begin{align}
    H_1 = \frac{1}{2c}\sum_i \mathbf{B}\cdot\mathbf{L}_i,
\end{align}
\begin{align}
    H_2 = \frac{1}{8c^2}\sum_i \left(\mathbf{B}\times \mathbf{r}_i \right)
    \cdot \left(\mathbf{B}\times \mathbf{r}_i \right),
\end{align}
where we have used the definition of the vector potential, and rewritten some terms using the electronic angular momentum operator, $\mathbf{L}_i=\mathbf{r}_i\times\mathbf{p}_i$. The derivative of the molecular energy with respect to the field is then calculated using the standard second-order perturbation theory, giving:
\begin{align}
\label{chipt}
\begin{split}
    \chi_{\mu\nu} &= -\langle 0| \frac{\partial^2 H_2}{\partial B_\mu \partial B_\nu}|0\rangle  \\
    &+\sum_{n\neq 0} \frac{\langle 0|\frac{\partial H_1}{\partial B_\mu}|n\rangle
    \langle n| \frac{\partial H_1}{\partial B_\nu}|0\rangle + 
    \langle 0|\frac{\partial H_1}{\partial B_\nu}|n\rangle
    \langle n| \frac{\partial H_1}{\partial B_\mu}|0\rangle}{E_n - E_0},
\end{split}
\end{align}
where $|n\rangle$ are the electronic wavefunctions of the system with energies $E_n$ ($n=0$ denotes the ground state). Note that the expectation value of the term $H_1$ is zero, because we assume that the molecule has closed-shell electronic structure. Therefore, it has no permanent magnetic moment and all terms linear in the field vanish.

The calculation of the derivatives with respect to the field strength in the above expression is elementary, and, after averaging over the orientations, we obtain the following formula for the isotropic magnetic susceptibility (in the zero-field limit):
\begin{align}
\label{chi0}
\begin{split}
    &\chi_0 = -\frac{1}{6c^2}\langle 0| r^2 |0\rangle 
    + \frac{1}{6c^2} \sum_{n\neq 0} \frac{|\langle 0|\mathbf{L}|n\rangle|^2}{E_n - E_0},
\end{split}
\end{align}
where we have introduced a shorthand notation $r^2=\sum_i (x_i^2 + y_i^2 + z_i^2)$ and $\mathbf{L}=\sum_i\mathbf{L}_i$ to simplify the formulas. As this expression is invariant to rotations, we are free to place the diatomic molecule along the $z$ axis. In this reference frame we have $L_z|0\rangle=0$, because the electronic ground state of the N$_2$ molecule is characterized by the $^1\Sigma_g^+$ molecular term. Additionally, $L_x|0\rangle = L_y|0\rangle$ due to the axial symmetry of the ground state wavefunction. This allows us to simplify the above formula to the form:
\begin{align}
\begin{split}
    &\chi_0 = -\frac{1}{6c^2}\langle 0| r^2 |0\rangle 
    + \frac{1}{3c^2} \sum_{n\neq 0} \frac{|\langle 0|L_x|n\rangle|^2}{E_n - E_0}.
\end{split}
\end{align}
The two terms in the above expression are frequently referred to as the diamagnetic and paramagnetic contributions, respectively, \cite{van1928dielectric} and we use this nomenclature in the following. Note that the paramagnetic term is strictly zero for atoms in the non-relativistic approximation when the gauge origin of the vector potential $\mathbf{A}$ is placed at the nucleus. The calculation of the diamagnetic term is straightforward, as it is just an expectation value of the multiplicative operator $r^2$ with the ground-state wavefunction. However, the paramagnetic term is more complicated as it formally involves a sum over all excited states, which is usually slowly convergent. As a result, a direct evaluation of this sum via determination of a sufficient number of excited states is computationally challenging. Therefore, we first rewrite the paramagnetic term in an equivalent form:
\begin{align}
\begin{split}
    \sum_{n\neq 0} \frac{|\langle 0|L_x|n\rangle|^2}{E_n - E_0} &= 
    \langle 0|L_x \sum_{n\neq 0} \frac{|n\rangle\langle n|}{E_n - E_0} L_x|0\rangle \\
    &=\langle 0|L_x \frac{Q}{H - E_0} L_x|0\rangle,
\end{split}
\end{align}
where we have used the closure relation to eliminate the sum over states; the projector $Q=1-|0\rangle\langle 0|$ is defined in the same way as in Eq.~(\ref{alpha0nuj2}). This formula enables us to use the coupled-cluster response theory to bypass the need for summation over the excited states, analogously to the definition of the polarizability in the limit $\omega\rightarrow 0$, see Eq.~(\ref{alpha0nuj2}). Indeed, let us define the response function $|\Psi_{\mathrm{resp}}\rangle$ as a solution of the equation:
\begin{align}
\label{resp}
    (H - E_0) |\Psi_{\mathrm{resp}}\rangle = QL_x|0\rangle,
\end{align}
then the magnetic susceptibility is given by the expression:
\begin{align}
    &\chi_0 = -\frac{1}{6c^2}\langle 0| r^2 |0\rangle 
    + \frac{1}{3c^2} \langle 0| L_x Q |\Psi_{\mathrm{resp}}\rangle.
\end{align}
Note that for a diatomic molecule, the diamagnetic and paramagnetic contributions to $\chi_0$, when treated separately, depend on the gauge origin of the vector potential $\mathbf{A}$. Only when these two terms are added together, the dependence on this parameter vanishes. However, this is strictly true when a complete basis set is used to represent the molecular orbitals, so any calculations involving a finite basis set would carry a residual gauge dependence. Of course, this effect becomes negligible if the results are saturated with respect to the basis set size. Nonetheless, for the sake of reproducibility, we state that the gauge origin was placed at the geometric center of the molecule in all calculations reported in this work.

Before moving on to the computational details, we remark that, in principle, the magnetic susceptibility of any molecule carries a frequency dependence, with full analogy with the dynamic polarizability. This was indicated in Eq.~(\ref{ll}) but was completely ignored in the formulation given above. Little is known about this dependence, as this effect has never been measured or calculated, and is usually assumed to be negligible. If one assumes conservatively that the frequency dependence of the magnetic susceptibility is of a similar magnitude as in the case of the polarizability, the effect will be of the order of 1--2\% at optical and microwave frequencies. However, there are reasons to believe that this effect is even smaller. In fact, the frequency dependence of $\chi_0$ can originate only from the paramagnetic term in Eq.~(\ref{chi0}), see the discussion in Refs.~\onlinecite{puchalski2023relativistic,lesiuk2024diamagnetic}. The diamagnetic contribution is not affected by the frequency of the external field. As we shall see shortly, the paramagnetic contribution to $\chi_0$ is roughly three times smaller than the diamagnetic term. Therefore, it is likely that the frequency dependence of $\chi_0$ is also smaller by a similar factor (in relative terms) in comparison with the polarizability. This suggests that the frequency dependence of the magnetic susceptibility is most likely smaller than 1\% at optical and microwave frequencies.

\begin{table*}
\caption{\label{tab:chi0}
Diamagnetic and paramagnetic contributions to the magnetic susceptibility of the nitrogen molecule, see Eq.~(\ref{chi0}), for three representative internuclear distances $R$, calculated using the frozen-core CC3 method within the aug-cc-pV$X$Z basis set family. All quantities are given in atomic units.
}
\begin{ruledtabular}
\begin{tabular}{lcccccc}
 $X$ & \multicolumn{2}{c}{$R=1.750$} & \multicolumn{2}{c}{$R=2.070$} & \multicolumn{2}{c}{$R=2.600$} \\
 \hline\\[-1.2em]
 & diamagnetic & paramagnetic & diamagnetic & paramagnetic & diamagnetic & paramagnetic \\
 \hline\\[-1.2em]
2 &  $-$0.000303 &   0.000110 &  $-$0.000350 &   0.000157 & $-$0.000436 &   0.000279  \\
3 &  $-$0.000300 &   0.000116 &  $-$0.000347 &   0.000168 & $-$0.000433 &   0.000300  \\
4 &  $-$0.000300 &   0.000117 &  $-$0.000346 &   0.000170 & $-$0.000432 &   0.000303  \\
5 &  $-$0.000299 &   0.000117 &  $-$0.000346 &   0.000170 & $-$0.000432 &   0.000304  \\
\end{tabular}
\end{ruledtabular}
\end{table*}

At a preliminary stage of the calculations, we found that the magnetic susceptibility of nitrogen molecule is about five orders of magnitude smaller than the polarizability. As only their sum enters the Lorentz--Lorenz equation, see Eq.~(\ref{ll}), that is relevant for the gas thermometry experiments, we arrive at the conclusion that the magnetic susceptibility does not have to be determined as accurately as the polarizability. In fact, the accuracy of about 10\% is acceptable from the experimental point of view, as this would translate to uncertainty less than one part per million in the total value of the coefficient that enters on the right-hand side of Eq.~(\ref{ll}). Taking this into account, we set our relative accuracy goal at 5\% or better for the magnetic susceptibility. If this level of precision is reached, the theoretical results can be used safely in the context of gas thermometry experiments, and the uncertainty of the calculated $\chi_0$ would bring a negligible contribution to the overall error budget.

Adoption of the 5\% accuracy target enables us to make some simplifications to the computational protocol used for $\chi_0$. First, we use the CC3 method for calculation of both the diamagnetic and paramagnetic contributions given in Eq.~(\ref{chi0}). This level of theory is also used for the coupled-cluster response function defined in Eq.~(\ref{resp}). We neglect all higher-order coupled cluster and relativistic corrections, cf. Eq.~(\ref{composite}). In the case of the static polarizability, the post-CC3 corrections constitute less than 1\% of the total value around the minimum of the potential, see Table~\ref{tab:ansummary}, so even if they turn out to be several times larger in the case of $\chi_0$, their omission is justified within the present accuracy target. The second simplification is the adoption of the frozen-core approximation with $1s^2$ core orbitals of both nitrogen atoms being uncorrelated. In the case of static polarizability, the contribution of core-core and mixed core-valence correlations is smaller than 0.35\% of the total value for all internuclear distances, see Table~\ref{tab:ansummary}, and it is unlikely that it becomes larger than 1\% in the case of magnetic susceptibility. Finally, we neglect the frequency dependence of $\chi_0$, see the discussion in the previous paragraph. Clearly, the combined effect of all approximations is extremely unlikely to exceed 5\%, even in the absence of any error cancellation.

In Table~\ref{tab:chi0}, we present the results of the calculations of the diamagnetic and paramagnetic contributions to the magnetic susceptibility for three representative internuclear distances ($R=1.750$~bohr, $R=2.070$~bohr, and $R=2.600$~bohr). We use the same aug-cc-pV$X$Z basis set family as in the case of polarizability. Somewhat surprisingly, both diamagnetic and paramagnetic terms converge extremely fast with the basis set size. In fact, from Table~\ref{tab:chi0} we see that the differences between the two largest basis sets ($X=4$ and $X=5$) are smaller than one part per thousand. The reason for this rapid convergence is the fact that the Hartree--Fock contribution to both diamagnetic and paramagnetic terms dominates, while the correlated contribution is relatively small. Therefore, there is little point in extending the basis set further or in performing extrapolation to the CBS limit. In the following, we take the results obtained with the $X=5$ basis set as the final estimate of $\chi_0$. In the Supplementary Material \cite{supp} we provide the calculated values of $\chi_0$ for the remaining internuclear distances, together with a plot illustrating the dependence on $R$ which is close to linear within the range of internuclear distances considered here.

\begin{table*}
\caption{\label{tab:fit}
Optimized linear and non-linear parameters of the fitting function in Eq.~(\ref{fitan}) for $\alpha_n(R)$ with $n=0,2,4,6$. All parameters are given in atomic units.}
\begin{ruledtabular}
\begin{tabular}{cd{8}d{8}d{8}d{8}}
 Param. & 
 \multicolumn{1}{c}{$n=0$} & 
 \multicolumn{1}{c}{$n=2$} & 
 \multicolumn{1}{c}{$n=4$} &
 \multicolumn{1}{c}{$n=6$} \\
 \hline\\[-1.2em]
 $c_{-1n}^{(1)}$ &   3973.239843 &   3121512.910661 &  2.9522897  &  19.85061337 \\
 $c_{0n}^{(1)}$  &  -2550.566901 &  -1565760.084872 & -1.06250446 &  -7.34704556 \\
 $c_{1n}^{(1)}$  &    480.40394  &    213071.693344 &  0.10234389 &   0.72912448 \\
 $c_{0n}^{(2)}$  & -17636.209308 & -12511573.39832  & -8.65155509 & -59.44131003 \\
 $c_{1n}^{(2)}$  &   6178.632412 &   4306441.860778 &  2.18269184 &  15.28369979 \\
 $c_{2n}^{(2)}$  &  -7984.197341 &  -4591578.708124 & -1.76012369 & -12.65835287 \\
 $a_n^{(1)}$     & 0.72712153 & 1.57852725 &  -6.04767072 & -5.57635061 \\
 $a_n^{(2)}$     & 3.06212416 & 3.56319374 &  -4.59332841 & -4.08555911 \\
\end{tabular}
\end{ruledtabular}
\end{table*}

As the final note, we point out that there is additional contribution to the magnetic susceptibility originating from the coupling of the molecular center-of-mass (COM) motion with the static external magnetic field, in full analogy with the atomic contributions analyzed by Bruch and Weinhold. \cite{bruch2002nuclear} Such effect is completely absent in the case of the static electric field which is not coupled to the COM motion (assuming that the molecule is chargeless). Moreover, some of these coupling terms depend on the temperature and thus bring an extra temperature dependence of the magnetic susceptibility outside of the standard rovibrational averaging described in Sec.~\ref{sec:temp}. Therefore, we have to analyze the potential impact of this effect and estimate its magnitude. To this end, we start with the total Hamiltonian of the molecule in the Born--Oppenheimer approximation including also the COM motion which separates from the electronic Hamiltonian in Eq.~(\ref{hel}) after application of the Power--Zienau--Woolley unitary transformation. \cite{johnson1983interaction} It reads:
\begin{align}
    H = \frac{\mathbf{K}^2}{2M_T} - \frac{1}{M_Tc}\,\mathbf{d}\cdot(\mathbf{K}\times\mathbf{B})
    +\frac{1}{2M_Tc^2}(\mathbf{d}\times\mathbf{B})^2,
\end{align}
where $M_T$ is the total mass of the molecule (nuclei plus electrons), $\mathbf{d}=-\mathbf{r}=-\sum_i \mathbf{r}_i$ is the total electronic dipole moment operator, and $\mathbf{K}$ is the pseudomomentum of the COM. As we are dealing with a macroscopic sample of a gas in temperatures where the quantum effects of the COM motion are negligible, we follow Bruch and Weinhold \cite{bruch2002nuclear} and adopt classical approximation for this degree of freedom, where $\mathbf{v}=\mathbf{K}/M_T$ is interpreted as a classical velocity of a pseudo-particle representing the COM motion. This enables us to simplify the Hamiltonian to the form:
\begin{align}
    H = \frac{1}{2}M_T \mathbf{v}^2 - \frac{1}{c}\,\mathbf{d}\cdot(\mathbf{v}\times\mathbf{B})
    +\frac{1}{2M_Tc^2}(\mathbf{d}\times\mathbf{B})^2,
\end{align}
where the first term is now simply the classical kinetic energy of COM. We also see that the second term can be interpreted as a classical Lorentz force acting upon the electronic cloud. For subsequent manipulations, it is useful to rewrite this term as $\mathbf{d}\cdot(\mathbf{v}\times\mathbf{B}) = \mathbf{B}\cdot(\mathbf{d}\times\mathbf{v})$, using the properties of the triple product to simplify the differentiation that follows. By applying the second-order perturbation theory to this Hamiltonian, see Eq.~(\ref{chipt}), we obtain COM coupling contribution to the isotropic magnetic susceptibility in the form:
\begin{align}
\label{chicom1}
\begin{split}
    &\chi_0^{\mathrm{COM}} =
    -\frac{2}{3M_Tc^2}\langle 0| \mathbf{d}^2 |0\rangle + 
    \frac{2}{3c^2}\sum_{n\neq 0}
    \frac{|\langle 0|(\mathbf{d}\times\mathbf{v})|n\rangle|^2}{E_n - E_0},
\end{split}
\end{align}
after averaging over the orientations. Because we are calculating susceptibility that is expected to be consistent with the measurements performed on a macroscopic sample of a gas, the COM velocity $\mathbf{v}$ is not a single number. Indeed, all quantities that contain $\mathbf{v}$ have to be averaged over the Maxwell--Boltzmann distribution at a given temperature. Using the three-dimensional Maxwell--Boltzmann distribution, the average values of $\mathrm{v}_\mu \mathrm{v}_\nu$ are given by the formula $\langle \mathrm{v}_\mu \mathrm{v}_\nu \rangle_{\mathrm{MB}} = \frac{kT}{M_T}\delta_{\mu\nu}$, where the subscript MB was added to avoid confusion with the quantum-mechanical bracket notation. Therefore, terms in Eq.~(\ref{chicom1}) that mix different Cartesian components of $\mathbf{v}$ bring no contribution to the thermal average. Taking into account this simplification we obtain:
\begin{align}
\label{chicom2}
\begin{split}
    &\chi_0^{\mathrm{COM}} =
    -\frac{2}{3M_Tc^2}
    \langle 0| \mathbf{d}^2 |0\rangle + 
    \frac{2\langle v_x^2 \rangle_{\mathrm{MB}}}{c^2}\cdot\frac{2}{3}
    \sum_{n\neq 0}\frac{|\langle 0|\mathbf{d}|n\rangle|^2}{E_n - E_0}.
\end{split}
\end{align}
We recognize that the last term (multiplied by the factor $2/3$) is just the static isotropic polarizability of the molecule, cf. Eq.~(\ref{alpha0nuj2}) with $k=0$, and insert the Maxwell--Boltzmann average $\langle \mathrm{v}_x^2 \rangle_{\mathrm{MB}}=kT/M_T$. This brings us to the final expression:
\begin{align}
\label{chicom3}
    &\chi_0^{\mathrm{COM}} =
    -\frac{2}{3M_Tc^2}
    \langle 0| \mathbf{d}^2 |0\rangle + 
    \frac{2kT}{M_Tc^2}\,\alpha_0.
\end{align}
Let us begin with estimating the magnitude of the first term of the above expression. The expectation value $\langle 0| \mathbf{d}^2 |0\rangle$ is not the same as $\langle 0| r^2 |0\rangle$ in Eq.~(\ref{chi0}), because the former involves additional two-electron terms such as $\langle \mathbf{r}_i\cdot\mathbf{r}_j\rangle$. Nonetheless, the order of magnitude of these two expectation values is expected to be roughly the same based on analogous data for noble gas atoms. Under this assumption, the first term of Eq.~(\ref{chicom3}) brings contribution of only around 0.01\% due to the enormous value of $M_T \approx 51\,000$ (in atomic units). Therefore, this correction is completely negligible in the present context. Moreover, if one insisted on taking this term into account, one would also need to consider analogous terms of the order $1/M_T$ that were neglected from the electronic Hamiltonian, see Eq.~(\ref{hel}), in the Born--Oppenheimer approximation.

The second term in Eq.~(\ref{chicom3}) is unique in the sense that it carries a temperature dependence on its own, unlike any other term, completely independently on the rovibrational structure of the molecule. As this term clearly increases with temperature, we consider the worst-case scenario of $T=2000\,$K which is the largest temperature considered in this work. Taking the static polarizability at $R=2.070$ from Table~\ref{tab:ansummary} and $kT\approx 0.0063$ in atomic units at $T=2000\,$K, we find that the second term in Eq.~(\ref{chicom3}) is even smaller than the first and contributes only few parts per million to the magnetic susceptibility. To sum up, the contributions to the magnetic susceptibility originating from the coupling of COM motion to the external magnetic field are negligible within the present accuracy requirements. Consequently, they are not taken into account in the following.

\section{Analytic representation of the results}
\label{sec:fitting}

To make the calculated data directly useful in the determination of the temperature dependence, a proper analytic representation of the static polarizability, Cauchy coefficients, and magnetic susceptibility must be provided. As we are dealing here with internuclear distances $R$ close to the equilibrium bond length, the fitting functions are purely short-range in nature, without the asymptotic terms involving inverse powers of $R$ considered in Paper~II in the case of the interaction potential. Consequently, the adopted analytic form used for the static polarizability and Cauchy coefficients reads:
\begin{align}
\label{fitan}
\begin{split}
    &\alpha_{n}(R) = 
    e^{-a_n^{(1)} R}\sum_{i=-1}^{1} c_{in}^{(1)}\,R^i + 
    e^{-a_n^{(2)} R}\sum_{i=0}^{2} c_{in}^{(2)}\,R^i,
\end{split}
\end{align}
while for the magnetic susceptibility a slightly simpler form is employed:
\begin{align}
\label{fitchi}
\begin{split}
    &\chi_{0}(R) = 
    e^{-b^{(1)} R}\sum_{i=0}^{1} d_{i}^{(1)}\,R^i + e^{-b^{(2)} R^2}\sum_{i=0}^{1} d_{i}^{(2)}\,R^i,
\end{split}
\end{align}
as the latter quantity does not have to be represented as accurately as the former. The linear parameters $c_{in}^{(1)}$, $c_{in}^{(2)}$, $d_i^{(1)}$, and $d_i^{(2)}$, as well the non-linear parameters $a_n^{(1)}$, $a_n^{(2)}$, $b^{(1)}$, and $b^{(2)}$ in the above expressions are found using least-squares approach analogous to the protocol used in Paper~II. Overall, the fitting procedure introduces negligible errors in the representation of the underlying quantities. On average, the error of the fit is by an order of magnitude smaller than the uncertainty of the data. For example, in the case of the static polarizability, the average error of the fit is roughly 0.01\% while the estimated uncertainty of the \emph{ab initio} data is roughly 0.1\%. Similarly, for the second Cauchy coefficient the corresponding values are 0.03\% and 0.23\%, respectively. The optimized parameters appearing in Eq.~(\ref{fitan}) are reported in Table~\ref{tab:fit}, while in the case of Eq.~(\ref{fitchi}) they are:  
$d_0^{(1)}=-0.0022528$, $d_1^{(1)}=0.0012629$,
$d_0^{(2)}=0.0012288$, $d_1^{(2)}=-0.0008785$,
$b^{(1)}=0.729884$, $b^{(2)}=0.195461$ (atomic units). Finally, we note that the fitting functions described above should be used only within the range of internuclear distances for which the theoretical calculations were carried out, i.e. $R\in[1.75,2.60]$, and are not expected to preserve a similar accuracy level when extrapolated outside of this interval.

In parallel to fitting the calculated \emph{ab initio} data, analytical representation of respective uncertainties shall be required later. In the case of the last Cauchy coefficient, $\alpha_6(R)$, and the magnetic susceptibility, we adopt a flat uncertainty of 10\% and 5\%, respectively, at each point, and therefore the uncertainties do not have to be represented in analytic form for these two quantities. For the remaining coefficients, $\alpha_0(R)$, $\alpha_2(R)$, and $\alpha_4(R)$, we employ the same analytic representation, Eq.~(\ref{fitan}), to describe the uncertainties. The uncertainties obtained from the fit for $\alpha_0(R)$ and $\alpha_2(R)$ are marginally larger (by about 5\% on average) than the uncertainties determined directly from the \emph{ab initio} data. Only in the case of $\alpha_4(R)$, this ratio is somewhat larger (15\% on average), but this is acceptable from a practical point of view and leads to only a slightly more conservative estimation of theoretical uncertainties. The optimized parameters of the fitting functions for the uncertainties of $\alpha_0(R)$, $\alpha_2(R)$, and $\alpha_4(R)$ are given in the Supplementary Material. \cite{supp}

\section{Final results and discussion}
\label{sec:lit}

\begin{figure}
\includegraphics[width=0.9\linewidth]{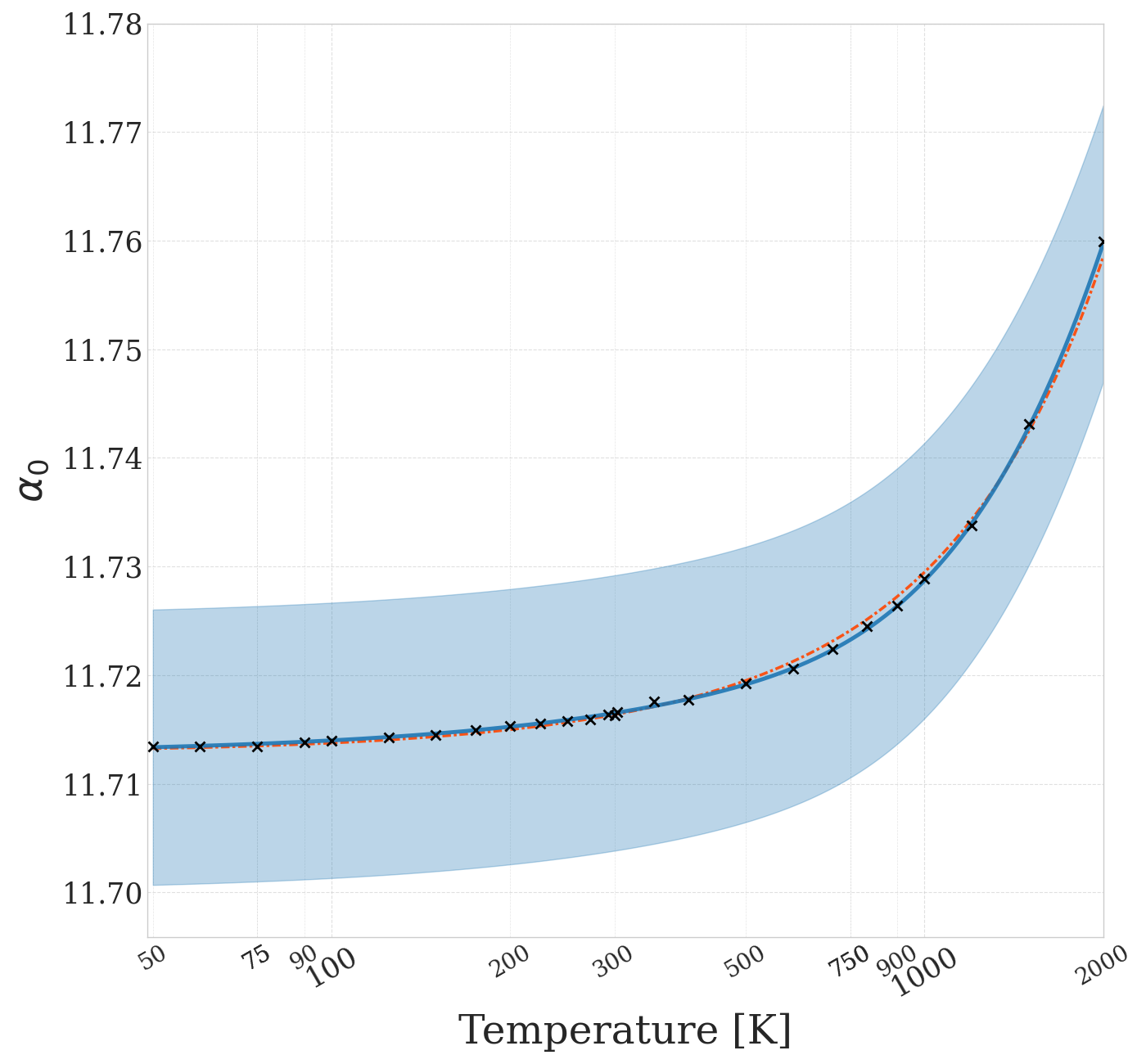}\\
\vspace{-0.1cm}
\includegraphics[width=0.9\linewidth]{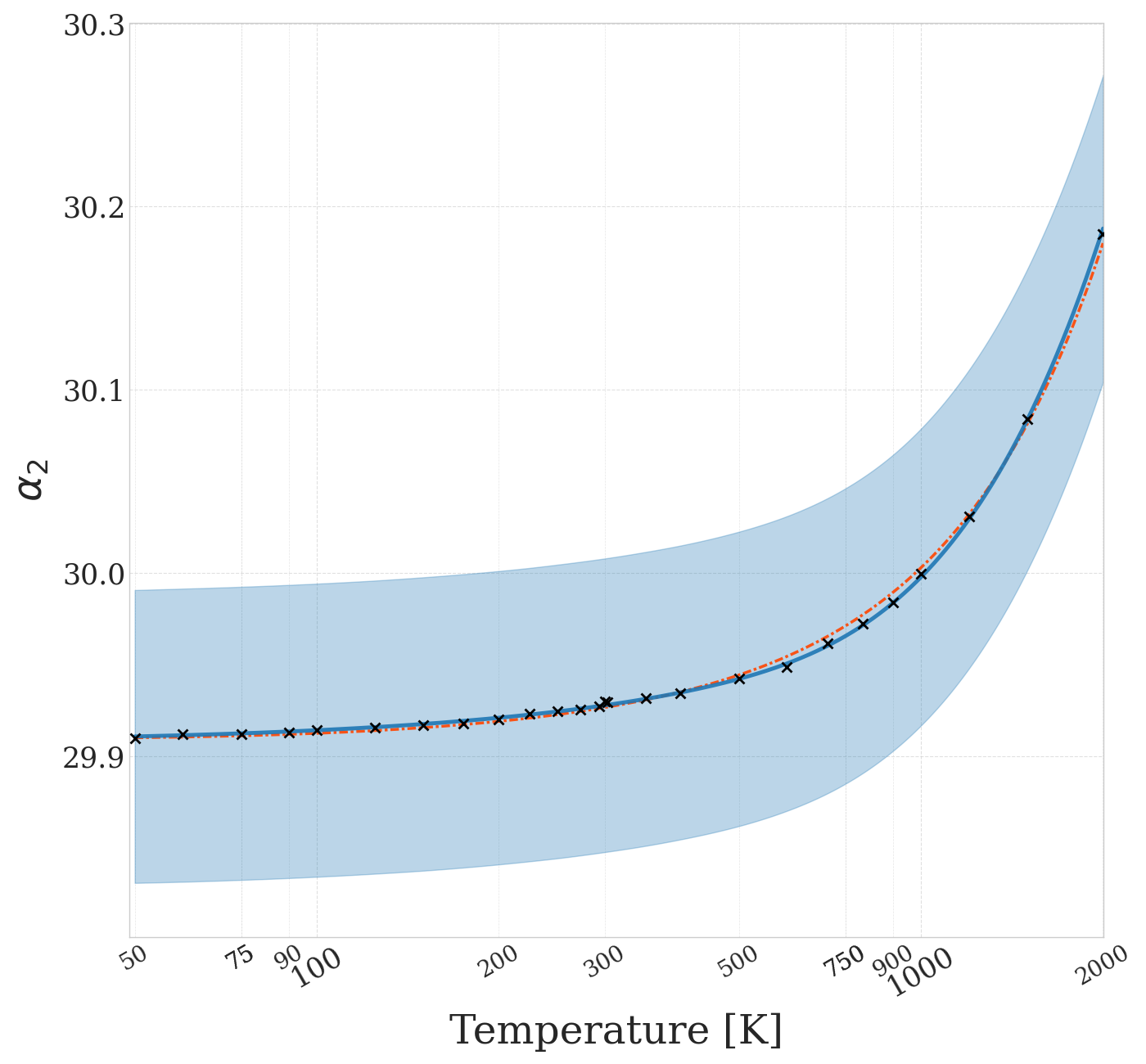}\\
\vspace{-0.1cm}
\includegraphics[width=0.9\linewidth]{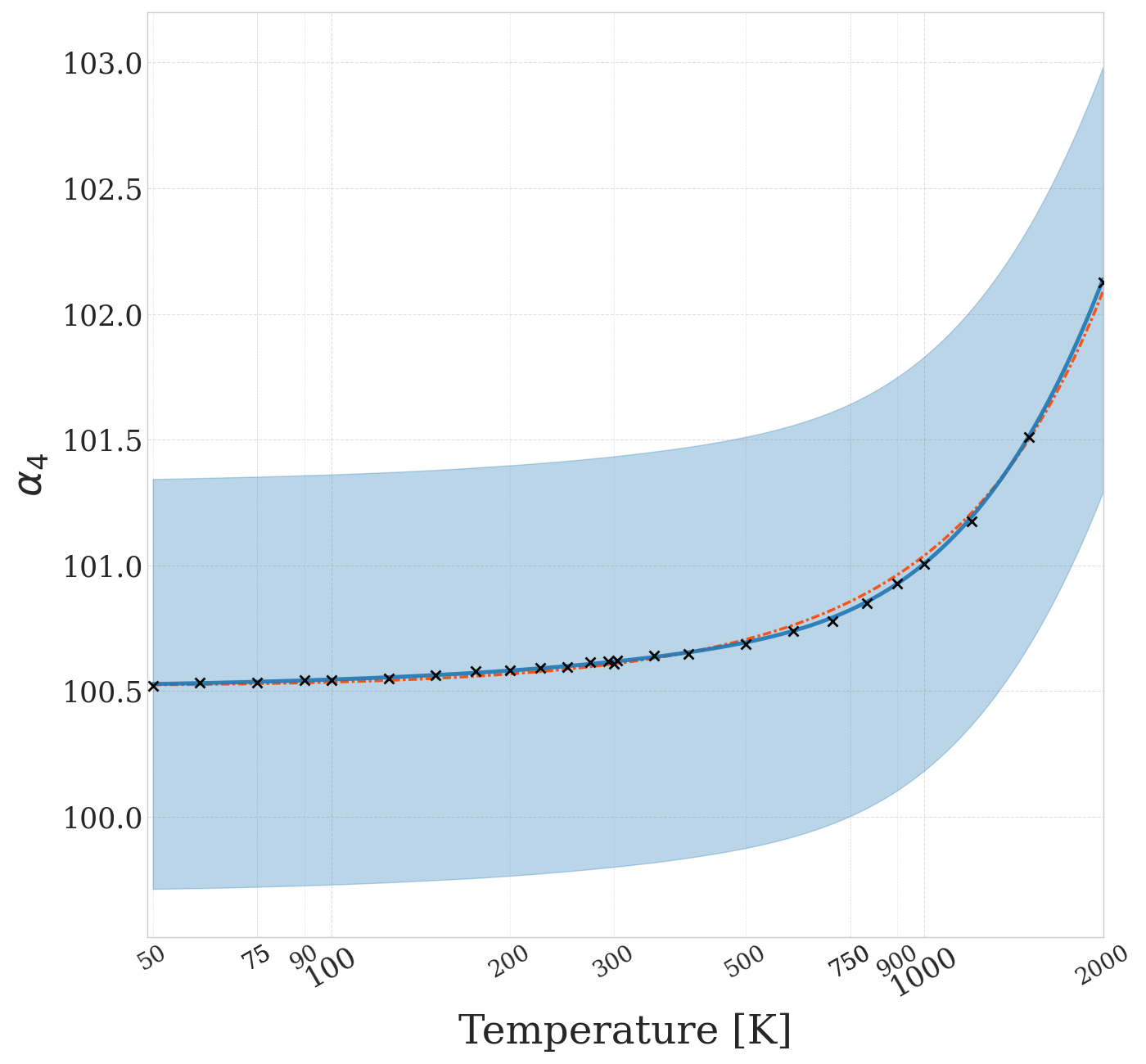}
\caption{Static polarizability (upper panel), second Cauchy coefficient (middle panel) and fourth Cauchy coefficient (bottom panel) for the nitrogen molecule as a function of temperature. The dark blue line represents results obtained using the rovibrational averaging method, see Sec.~\ref{sec:rovibro} (interpolated for clarity). The black crosses are results obtained with the PIMC approach, see Sec.~\ref{sec:pimc}. The light-blue shaded area are the total uncertainty estimates of the calculated quantities. The dashed orange curves are the fitting functions, Eq.~(\ref{eq:tcorrelation}).
The temperature is given in kelvin and the remaining quantities in atomic units.
\label{fig:polar}
}
\end{figure}

\begin{figure}
\includegraphics[width=0.9\linewidth]{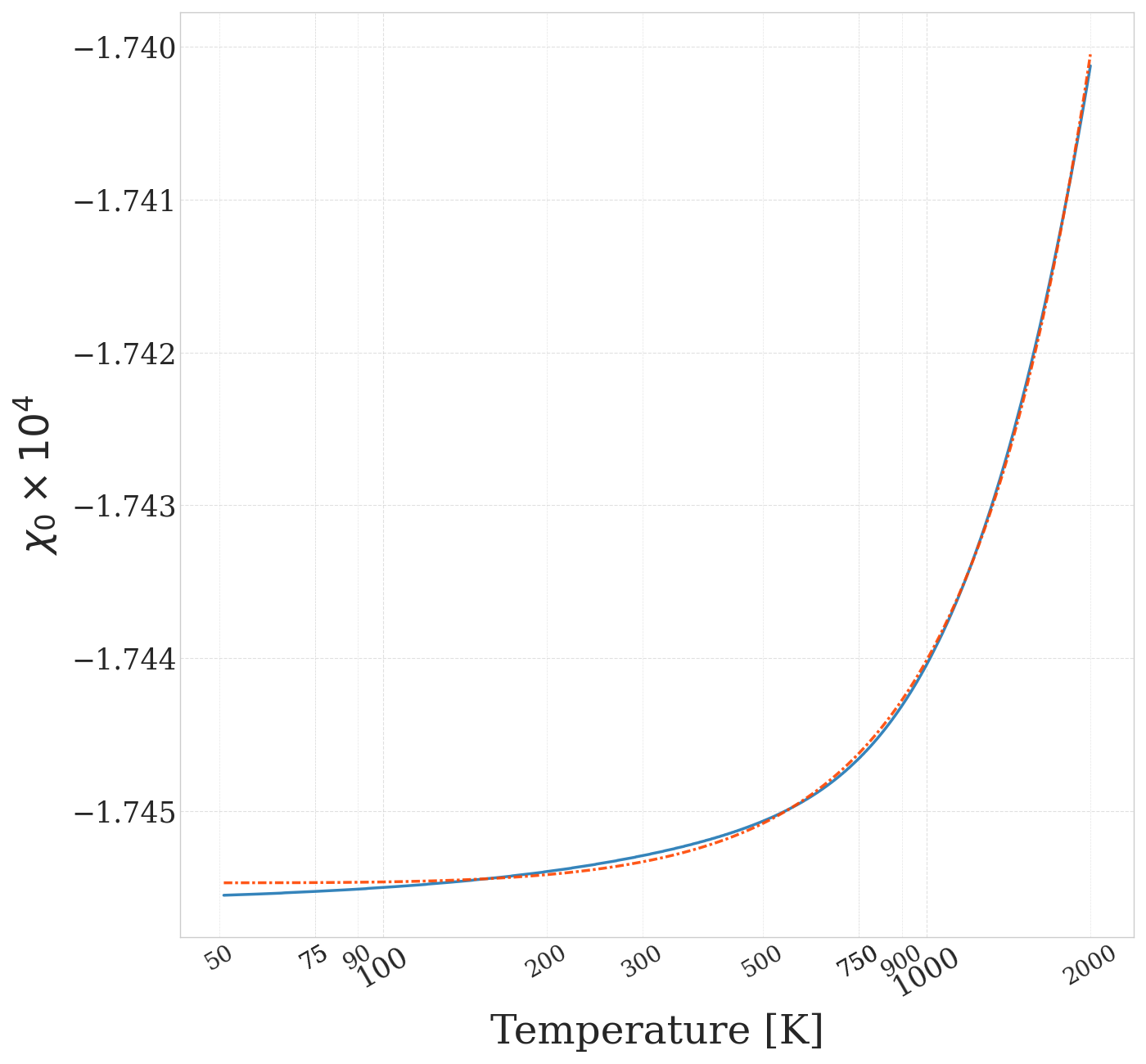}
\caption{Magnetic susceptibility of the nitrogen molecule as a function of temperature. The results were obtained using the rovibrational averaging method, see Sec.~\ref{sec:rovibro} (interpolated for clarity). The dashed orange curve is the fitting function, Eq.~(\ref{eq:tcorrelation}). Error bars are omitted for visual clarity, as their width obscures the upward trend of $\chi(T)$. The temperature is given in kelvin and the magnetic susceptibility in atomic units.}
\label{fig:chi0}
\end{figure}

In Fig.~\ref{fig:polar} we present the final results of the calculations of the polarizability $\alpha_0(T)$ and Cauchy coefficients, $\alpha_2(T)$ and $\alpha_4(T)$, for the nitrogen molecule as a function of temperature. The corresponding plot for the Cauchy coefficient $\alpha_6(T)$ is given in Supplementary Material. The temperature dependence was incorporated using two methods: rovibrational averaging and PIMC, as described in Sec.~\ref{sec:rovibro} and Sec.~\ref{sec:pimc}, respectively.  The rovibrational averaging was carried out for temperatures from $T=50\,$K to $T=2000\,$K in steps of $1\,$K. Due to the higher cost of the PIMC calculations, a sparser grid of temperatures was employed, but with a sufficiently large number of points to confirm the results of the former method. As can be seen from Fig.~\ref{fig:polar}, a near-perfect agreement is observed between the two methods. The differences are within the statistical uncertainty of the PIMC calculation, which is at least two orders of magnitude smaller than the inherent error of \emph{ab initio} calculations---propagated using Eq.~(\ref{eq:U_total})---and therefore negligible in comparison. In Fig.~\ref{fig:polar} we additionally present the uncertainty of the calculated polarizabilities and Cauchy coefficients evaluated as described in Sec.~\ref{sec:pimc}, and interpolated to match the temperatures used in the rovibrational averaging method.

In some applications, it may be more convenient to use the temperature dependence of the polarizability and the Cauchy coefficients represented as an analytic function rather than on a grid of temperatures. To address such a situation, we fitted the functions $\alpha_n(T)$ within the whole temperature range using the following formula:
\begin{align}
\label{eq:tcorrelation}
    \alpha_n(T) = A_0 + 
    \frac{c_1 + c_2 T}{e^{\theta/T} - 1},
\end{align}
where $A_0$, $c_1$, $c_2$, and $\theta$ are adjustable parameters. The values of these parameters for each $n$ are included in the Supplementary Material. For illustration, the fitting functions~ are also included in Fig.~\ref{fig:polar}. Clearly, the accuracy of Eq.~(\ref{eq:tcorrelation}) is entirely sufficient and the fit deviates from the explicitly calculated data by less than the inherent uncertainty of the latter.

In Fig.~\ref{fig:chi0} we plot the magnetic susceptibility of nitrogen molecule as a function of temperature. Clearly, this dependence for $\chi_0$ is much weaker than in the case of the polarizability and Cauchy coefficients. The complete set of data generated by the calculations described here is given in the Supplementary Material. The temperature dependence of the magnetic susceptibility has also been represented using the analytic form~(\ref{eq:tcorrelation}) and the corresponding coefficients are given in the Supplemental Material.

For the purposes of further discussion, we explicitly write here the results (in atomic units) for $T=303\,$K which is the same temperature as in the most recent experimental studies:
\begin{align}
\label{data303}
\begin{split}
    &\alpha_0 = 11.717(13), \\
    &\alpha_2 = 29.928(80), \\
    &\alpha_4 = 100.62(82), \\
    &\alpha_6 = 383(38), \\
    &\chi_0 = -1.75(9)\cdot 10^{-4}.
\end{split}
\end{align}
The uncertainties of the first three quantities (evaluated at $2\sigma$ confidence level) include the propagated uncertainty from the potential and the Cauchy moments, computed using PIMC according to Eq. (\ref{eq:U_total}). The statistical uncertainty of the PIMC calculation itself is negligible. The uncertainties of $\alpha_6$ and $\chi_0$ coefficients are set at the level of 10\% and 5\%, respectively, see the discussion in Sec.~\ref{subsec:alpha} and Sec.~\ref{subsec:chi0}.

The polarizability of the nitrogen molecule has been the subject of numerous previous theoretical studies, see Refs.~\onlinecite{luo1993frequency,sauer1994correlated,hattig1996tdmp2,sekino1993molecular,hammond2008coupled,neiss2007frequency,monten2011many,loukhovitski2016influence,jaszunski2001coupled,spelsberg1999dynamic,christiansen1998polarizabilities} and references therein,
and it was also determined semi-empirically with the dipole oscillator strength distribution (DOSD) technique, see the article of Kumar and Meath. \cite{kumar1992constrained} However, the most reliable data available in the literature for the polarizability of nitrogen molecule come from refractivity experiments and in the following we compare our results directly with these measurements. The most recent determinations by Egan and Yang \cite{egan2023optical,egan2024optical,yang2025demonstration} were focused on the molar refractivity coefficient defined as:
\begin{align}
    A_R(\omega,T) = \frac{4\pi N_A}{3} \left[ \alpha(\omega,T) + \chi(\omega,T) \right],
\end{align}
which enters the right-hand side of Eq.~(\ref{ll}). Neglecting the frequency dependence of the magnetic susceptibility and expanding the dynamic polarizability in the Cauchy series we find:
\begin{align}
\label{arexp}
\begin{split}
    &A_R(\omega,T) = \frac{4\pi N_A}{3} \Big[ 
    \alpha_0(T) + \alpha_2(T)\,\omega^2 \\
    &+ \alpha_4(T)\,\omega^4 + 
    \alpha_6(T)\,\omega^6 + \chi_0(T) \Big].
\end{split}
\end{align}
This enables us to calculate the molar polarizability $\alpha(\omega, T)$ using the theoretical data shown in Eq.~(\ref{data303}) corresponding to the temperature $T = 303\,$K, the same as the measurements of Egan and Yang. We obtained $11.743(13)$ and $11.875(13)$ for the wavelengths of the electric field $1542.383\,$nm and $632.9908\,$nm, respectively, at $2\sigma$ uncertainty level. The corresponding experimental results read $11.762\,120(37)$ and $11.894\,428(37)$. We see that the experimental data is slightly outside the predicted theoretical error bars at $2\sigma$ uncertainty level, but are well within the $3\sigma$ error bars. It is also worth pointing out that the error bars reported by Egan and Yang are given at $1\sigma$ uncertainty level (standard uncertainties).

From the above analysis it is clear that the accuracy of the theoretical results for the polarizability is worse by roughly two orders of magnitude than that of the experimental data. Therefore, theoretical calculations cannot be used as a replacement for direct measurements at present. However, this is not the main goal of the present study. The main purpose of the theoretical results is, for example, to enable determination of the molar polarizability at different wavelengths or temperatures where no measurements are available, or to calculate quantities that have not been directly measured thus far. As an example of the latter approach, we combine the theoretical and experimental data to derive the best estimate of the static polarizability of the nitrogen molecule. This quantity has not been measured thus far with accuracy comparable to the work of Egan and Yang. To fill this gap, we combine two experimental results of Egan and Yang at wavelengths $1542.383\,$nm and $632.9908\,$nm with the coefficients $\alpha_4(T)$, $\alpha_6(T)$ and $\chi_0(T)$ obtained from theory. Returning to Eq.~(\ref{arexp}), this enables us to write a pair of equations corresponding to two wavelengths and solve them to get the value of $\alpha_0(T)$. By following this procedure, we obtained $\alpha_0(T)=11.735\,962$~a.u. for the static polarizability of nitrogen molecule at $T=303\,$K. The accuracy of this result is expected to be no worse than $10\,$ppm, taking into account the combined uncertainties of the measurements and of the theoretical data. Note that this value differs somewhat from the estimate of Egan and Yang, $11.735\,909$~a.u., who used an identical procedure to extract the polarizability, but with older experimental or theoretical data for $\alpha_4(T)$, $\alpha_6(T)$ and $\chi_0(T)$. We believe that the value of $\alpha_0(T)$ derived above can serve as a useful reference point for future calculations or experiments. Additionally, using the temperature dependence of the static polarizability calculated in this work, this property can be determined at any other temperature that one is interested in. For example, by adding the difference in the polarizabilities obtained from the theory at $T=303\,$K and $T=273.16\,$K, see the Supplementary Material, \cite{supp} we find $\alpha_0(T)=11.735\,585$~a.u. at $T=273.16\,$K. This temperature (the triple point of water) is particularly important as one of the frequently used reference temperatures from the international temperature standard (ITS-90). \cite{preston1990international} Note that as the additional result of the procedure described above, we also obtain a semi-empirical estimate of the second Cauchy coefficient $\alpha_2(T)$. This gives $30.086\,359$~a.u at $T=303\,$K which differs somewhat from the value obtained by Egan and collaborators \cite{yang2025demonstration}, $30.109$~a.u. However, this increased deviation is understandable, considering that $\alpha_2(T)$ is more sensitive to the values of the higher-order coefficients used in the procedure which are different here than in the paper of Egan and Yang, see the discussion above.

Finally, regarding the magnetic susceptibility, the data available in the literature is particularly scarce. We are not aware of any recent rigorous calculation of this quantity, aside from an unpublished dataset from our group, where the paramagnetic part of the magnetic susceptibility was neglected and a counter-term was added \emph{ad hoc} to the diamagnetic part to eliminate the gauge dependence. However, as is evident from this work, the paramagnetic part is substantial and in fact cannot be neglected even in a crude approximation scheme. Therefore, we strongly recommend using the theoretical data reported in this work instead of the previous unpublished results. Simultaneously, all measurements of this quantity, to the best of our knowledge, date back roughly 100 years. The first is the work of Hector \cite{hector24}, which is also cited in the commonly used \emph{CRC Handbook of Chemistry and Physics} \cite{noauthororeditor2007handbook}, reporting $\chi_0=-1.34\cdot10^{-4}$ (in atomic units) and stating the uncertainty at 2\%. Next, we have the work of Vaidyanathan \cite{vaidyanathan1928}, giving $\chi_0=-1.45\cdot10^{-4}$ at $T=303\,$K and estimating the error at no more than 5\%. Next, we have the work of Bitter~\cite{bitter1930} who gives $\chi_0=-1.66\cdot10^{-4}$ at $T=298\,$K but without explicit uncertainty estimation. Finally, we have the measurement by Havens \cite{havens1933}, yielding $\chi_0=-1.338(2)\cdot10^{-4}$ at room temperature. There are also earlier measurements by Son\'e \cite{sone1920} and Hammar \cite{hammar1926}, which give results roughly two times smaller than the data cited above, which is usually attributed to the issues with gas purity \cite{havens1933}.

It is clear that the theoretical value, $\chi_0 = -1.75(9)\cdot 10^{-4}$, determined by us at $T=303\,$K differs substantially from the aforementioned experimental results. We cannot explain the origin of this discrepancy at present, even after a careful analysis, see Sec.~\ref{subsec:chi0}. We believe that it is extremely unlikely that contributions to $\chi_0$ that were neglected in this work, i.e., frequency dependence, mass dependence, or relativistic effects, can explain such a large difference. Therefore, we conclude that the discrepancy is due to either a fundamental problem with the adopted theoretical framework or a measurement or interpretation error. Finally, we note that a new experiment or independent calculations by a different group, or both, are needed to explain the discrepancy. This is especially true if one takes into consideration that a similar discrepancy between theory and experiment persists also for noble gases.\cite{lesiuk20,lesiuk2024diamagnetic,lesiuk23,puchalski2023relativistic}

\section{Conclusions}
\label{sec:concl}

In this work, we have presented a comprehensive theoretical determination of the temperature dependence of the molecular polarizability and magnetic susceptibility of the nitrogen molecule, as well as of the frequency dependence of the former quantity. Motivated by the stringent accuracy requirements of modern gas thermometry experiments and precision metrology in general, we employed a state-of-the-art composite coupled-cluster approach to calculate the purely electronic contributions to the static polarizability, Cauchy coefficients, and isotropic magnetic susceptibility across a relevant grid of internuclear distances. The integration of quantum nuclear effects and temperature dependence was achieved through two independent theoretical frameworks: rigorous rovibrational averaging and the path integral Monte Carlo (PIMC) method. The near-perfect agreement between these two approaches across the $50-2000\,$K range confirms the numerical stability and methodological robustness of our treatment of the thermal effects. 

By comparing our theoretical results with recent high-precision RIGT measurements, we confirmed that our calculations agree with the experimental data whenever the latter are available. Equally importantly, we demonstrated the practical utility of our theoretical results by adopting a hybrid approach. By combining the highly accurate experimental molar polarizability data with our theoretically derived higher-order Cauchy coefficients and magnetic susceptibility, we obtained new highly-accurate semi-empirical estimates for the static polarizability of N$_2$ at standard reference temperatures, such as $T=303\,$K and the water triple point $T=273.16\,$K. 

Finally, our investigation into the magnetic susceptibility of N$_2$ revealed that the paramagnetic contribution is substantial and cannot be omitted. We demonstrated that secondary effects, such as the coupling of the center-of-mass motion to the external magnetic field, are negligible within our target accuracy. Notably, our theoretical value diverges significantly from the single experimental result available in the literature. Given that omitted theoretical corrections (such as frequency dependence and higher-order relativistic effects) are far too small to bridge this gap, the reason for this discrepancy is not known at present.

\section*{Supplementary Material}

See the Supplementary Material for the list of frequencies and internuclear distances considered in this work, raw results of all calculations reported in the main text, parameters of the optimized fitting functions of the uncertainties, list of temperature-dependent data for static polarizability, Cauchy coefficients, and magnetic susceptibility, and Python implementation of all fitting functions used in this work.

\begin{acknowledgments}
The project (22IEM04 MQB-Pascal) has received funding from the European Partnership on Metrology, co-financed from the European Union’s Horizon Europe Research and Innovation Programme and by the Participating States. We gratefully acknowledge Poland's high-performance Infrastructure PLGrid (HPC Centers: ACK Cyfronet AGH, PCSS, CI TASK, WCSS) for providing computer facilities and support within computational grant PLG/2025/018692. The authors also thank Pozna\'n Supercomputing and Networking Center for the computational grant pl0458-01. We are grateful to Dr. Allan H. Harvey (NIST) and Dr. Christian G\"unz (PTB) for fruitful discussions as well as their insightful comments on the paper.
\end{acknowledgments}

\section*{Data Availability Statement}

The data that support the findings of this study are available within the article and its supplementary material, and openly available in Zenodo repository at \url{https://doi.org/10.5281/zenodo.22662189}, reference number 22662189.

\bibliography{nitrogen-paper3}

\end{document}